%% file: paper.tex
\documentclass[twocolumn]{aastex631}
\usepackage{amsmath}

\newcommand{\citepe}[1]{\citep[e.g.,][]{#1}}

\newcommand{\comment}[1]{}
\newcommand{\Add}[1]{\textcolor{black}{#1}}
\DeclareRobustCommand{\Erase}{\bgroup\markoverwith{\textcolor{red}{\rule[.5ex]{2pt}{0.4pt}}}\ULon}

\newcommand{\M}{{\ $\mid$}}

\newcommand{\W}{{$\lambda$}}

\newcommand{\OII}{[O\,\textsc{ii}]}

\newcommand{\Ha}{H$\alpha$}
\newcommand{\Hb}{H$\beta$}
\newcommand{\Hc}{H$\gamma$}
\newcommand{\Hd}{H$\delta$}
\newcommand{\CIIIsemi}{C\,\textsc{iii}]}
\newcommand{\CIV}{C\,\textsc{iv}}
\newcommand{\NII}{[N\,\textsc{ii}]}

\newcommand{\OIII}{[O\,\textsc{iii}]}

\newcommand{\SII}{[S\,\textsc{ii}]}

\newcommand{\HeII}{He\,\textsc{ii}}
\newcommand{\HeI}{He\,\textsc{i}}

\newcommand{\NeIII}{[Ne\,\textsc{iii}]}
\newcommand{\NeV}{[Ne\,\textsc{v}]}

\newcommand{\ergs}{erg\,s$^{-1}$}
\newcommand{\ergscm}{erg\,s$^{-1}$\,cm$^{-2}$}

\newcommand{\kms}{km\,s$^{-1}$}
\newcommand{\metal}{$12$+$\log ({\rm O/H})$}

\newcommand{\HII}{H\,\textsc{ii}}

\newcommand{\EW}{$\textrm{EW}_0$}

\newcommand{\q}{$q_{\rm ion}$}

\newcommand{\EMPGa}{J1631+4426} 
\newcommand{\EMPGb}{J104457} 
\newcommand{\EMPGc}{I Zw 18 NW} 

\shorttitle{Ionizing Spectrum Shapes of Young Galaxies}
\shortauthors{Umeda et al.}
\graphicspath{{./}{figures/}}

\begin{document}
\defcitealias{2021ApJ...913...22K}{K21}
\defcitealias{2021arXiv210512765B}{B21}
\defcitealias{2005ApJS..161..240T}{TI05}
\defcitealias{2021arXiv210906725O}{O21}

\title{EMPRESS. VII. \\
Ionizing Spectrum Shapes of Extremely Metal-Poor Galaxies:\\
Uncovering the Origins of Strong He{\sc ii} and the Impact on Cosmic Reionization}

\author{Hiroya Umeda}
\affiliation{Institute for Cosmic Ray Research,
The University of Tokyo,
5-1-5 Kashiwanoha, Kashiwa,
Chiba 277-8582, Japan; \email{ume@icrr.u-tokyo.ac.jp}}
\affiliation{Department of Physics, Graduate School of Science, The University of Tokyo, 7-3-1 Hongo, Bunkyo, Tokyo 113-0033, Japan}

\author[0000-0002-1049-6658]{Masami Ouchi}
\affiliation{National Astronomical Observatory of Japan, 2-21-1 Osawa, Mitaka, Tokyo 181-8588, Japan}
\affiliation{Institute for Cosmic Ray Research,
The University of Tokyo,
5-1-5 Kashiwanoha, Kashiwa,
Chiba 277-8582, Japan; \email{ume@icrr.u-tokyo.ac.jp}}
\affiliation{Kavli Institute for the Physics and Mathematics of the Universe (WPI), 
University of Tokyo, Kashiwa, Chiba 277-8583, Japan}

\author[0000-0003-2965-5070]{Kimihiko Nakajima}
\affiliation{National Astronomical Observatory of Japan, 2-21-1 Osawa, Mitaka, Tokyo 181-8588, Japan}

\author[0000-0001-7730-8634]{Yuki Isobe}
\affiliation{Institute for Cosmic Ray Research,
The University of Tokyo,
5-1-5 Kashiwanoha, Kashiwa,
Chiba 277-8582, Japan; \email{ume@icrr.u-tokyo.ac.jp}}
\affiliation{Department of Physics, Graduate School of Science, The University of Tokyo, 7-3-1 Hongo, Bunkyo, Tokyo 113-0033, Japan}

\author[0000-0002-1005-4120]{Shohei Aoyama}
\affiliation{Institute for Cosmic Ray Research,
The University of Tokyo,
5-1-5 Kashiwanoha, Kashiwa,
Chiba 277-8582, Japan; \email{ume@icrr.u-tokyo.ac.jp}}

\author[0000-0002-6047-430X]{Yuichi Harikane}
\affiliation{Institute for Cosmic Ray Research,
The University of Tokyo,
5-1-5 Kashiwanoha, Kashiwa,
Chiba 277-8582, Japan; \email{ume@icrr.u-tokyo.ac.jp}}
\affiliation{Department of Physics and Astronomy, University College London, Gower Street, London WC1E 6BT, UK}

\author[0000-0001-9011-7605]{Yoshiaki Ono}
\affiliation{Institute for Cosmic Ray Research,
The University of Tokyo,
5-1-5 Kashiwanoha, Kashiwa,
Chiba 277-8582, Japan; \email{ume@icrr.u-tokyo.ac.jp}}

\author{Akinori Matsumoto}
\affiliation{Institute for Cosmic Ray Research,
The University of Tokyo,
5-1-5 Kashiwanoha, Kashiwa,
Chiba 277-8582, Japan; \email{ume@icrr.u-tokyo.ac.jp}}
\affiliation{Department of Physics, Graduate School of Science, The University of Tokyo, 7-3-1 Hongo, Bunkyo, Tokyo 113-0033, Japan}


\begin{abstract}

Strong high-ionization lines such as {\HeII} of young galaxies are puzzling at high and low redshift. Although recent studies suggest the existence of non-thermal sources, whether their ionizing spectra can consistently explain multiple major emission lines remains a question. Here we derive the general shapes of the ionizing spectra for three local extremely metal-poor galaxies (EMPGs) that show strong {\HeII}$\lambda$4686.
We parameterize the ionizing spectra composed of a blackbody and power-law radiation mimicking various stellar and non-thermal sources.
We use photoionization models for nebulae,
and determine seven parameters of the ionizing spectra and nebulae by Markov Chain Monte Carlo methods, carefully avoiding systematics of abundance ratios. 
We obtain the general shapes of ionizing spectra explaining $\sim 10$ major emission lines within observational errors with smooth connections from observed X-ray and optical continua.
We find that an ionizing spectrum of one EMPG has a blackbody-dominated shape, while the others have convex downward shapes at $>13.6$ eV, which indicate a diversity of the ionizing spectrum shapes. We confirm that the convex downward shapes are fundamentally different from ordinary stellar spectrum shapes, and  that the spectrum shapes of these galaxies are generally explained by the combination of the stellar and ultra-luminous X-ray sources. Comparisons with stellar synthesis models suggest that the diversity of the spectrum shapes arises from differences in the stellar age. If galaxies at $z\gtrsim 6$ are similar to the EMPGs, high energy ($>54.4$ eV) photons of the non-stellar sources negligibly contribute to cosmic reionization due to relatively weak radiation.

\end{abstract}


\section{Introduction} \label{sec:intro}
Understanding the epoch of reionization (EoR) is one of the most important goals for astronomy today. Many studies have suggested that galaxies are the dominant sources of ionizing photons at EoR \citepe{2014MNRAS.442.2560W,2015ApJ...813L...8M,2016MNRAS.456..485S}. 
From the hydrodynamical simulations, high-redshift young galaxies at the early formation phase are expected to have low metallicities, low stellar masses, and high specific star formation rates \citepe{2012ApJ...745...50W}. 
High-redshift galaxies are studied with various large telescopes including the {\it Hubble Space Telescope} ({\it HST}) \citepe{2015ApJ...811..140B,2017ApJ...835..113L,2018MNRAS.479.5184A,2018ApJ...855..105O,2016ApJ...821..123H,2020ApJ...893...60K}. 
Although there are spectroscopic studies for very bright or lensed galaxies at high redshifts \citepe{2018PASJ...70S..15S,2018MNRAS.479.1180M,2015MNRAS.450.1846S}, dwarf galaxies at high-redshift are too faint to be detected with current observational instruments. For example, even with upcoming {\it James Webb Space Telescope} ({\it JWST}), we can detect {\Ha} for galaxies  with stellar mass below $10^6~M_\odot$ only at $z\lesssim2$ without gravitational lensing \citep{2021arXiv210803850I}\par
Owing to the observational difficulties, many studies use local dwarf galaxies as analogs of high-redshift young dwarf galaxies \citepe{2019ApJ...878L...3B}. Among the local dwarf galaxies, local extremely metal-poor galaxies (EMPGs) are recently gaining attention \citepe{1998ApJ...497..227I,2005ApJS..161..240T,2016ApJ...819..110S,2020ApJ...898..142K}. 
EMPGs are defined as galaxies with metallicity below $10\%$ solar metallicity $Z_\odot$. So far, EMPGs with metallicities down to $1.6\%~Z_{\odot}$ have been identified using data obtained with Subaru/Hyper Suprime Cam (HSC) \citep{2020ApJ...898..142K}. Local EMPGs have low masses and high specific star-formation rates (sSFRs), both of which are comparable to those of high-redshift young galaxies \citepe{2012ApJ...745...50W}. Local EMPGs are therefore regarded as local analogs of high-redshift young galaxies.\par
The importance of relationships between local EMPGs and high-redshift galaxies is further emphasized because very strong high-ionization nebular emission lines (i.e., {\HeII}, {\CIIIsemi}, {\CIV}) are detected from local EMPGs \citep{2019ApJ...878L...3B,2017MNRAS.472.2608S,2000ApJ...531..776G,2012MNRAS.421.1043S,2021ApJ...913...22K} 
and high-redshift galaxies \citep{2015MNRAS.450.1846S,2017ApJ...839...17S,2017ApJ...851...40L,2018MNRAS.479.1180M}. 
Most of the EMPGs have detections of optical recombination lines {\HeII\W4686} that need high energy ionizing photons ($>54.4~{\rm eV}$) \citep{2010A&A...516A.104L,2012MNRAS.421.1043S,2017MNRAS.472.2608S}. A statistical study of local galaxies shows that fluxes of {\HeII~$\lambda4686$} increase as the metallicities decrease \citepe{2019A&A...622L..10S}. These results indicate the presence of hard ionizing sources in the metal-poor environments.\par
The origin of strong {\HeII} emission line is still under debate. \cite{2018MNRAS.477..904X} examine contributions from stellar radiation using the photoionization code {\sc cloudy} \citep{2013RMxAA..49..137F} and the Binary Population and Spectral Synthesis (BPASS) code \citep{2016MNRAS.456..485S,2017PASA...34...58E}. BPASS binary models incorporate contributions from hard radiation from Wolf-Rayet stars and stripped binary stars. However, a line ratio {\HeII\W4686}/{\Hb} predicted in \cite{2018MNRAS.477..904X} falls short of the observed value by more than 0.5 dex, suggesting that stars are not major contributors to the {\HeII} ionizing photons. Addition of extra stars (e.g., metal-poor, very massive, or Wolf-Rayet stars) could boost the {\HeII} ionizing photons, but several studies suggest that these stars cannot fully account for the observed strength of {\HeII} in EMPGs \citepe{2018MNRAS.480.1081K,2021arXiv210906725O}. \Add{\cite{2020A&A...643A..80P} suggest that ordinary stellar synthesis models with non-zero ionizing photon escape fraction can explain {\HeII\W4686}/{\Hb}, but their models require very high escape fraction values ($\sim70\%$) that are not compatible with the typical values reported for $z\lesssim 3$ galaxies \citepe{2021MNRAS.503.1734I,2016ApJ...819...81R,2020ApJ...904...59A,2018ApJ...869..123S,2022MNRAS.511..120S}.}\par
Non-thermal ionizing sources have been proposed to explain the strong nebular {\HeII} from EMPGs. Active galactic nuclei (AGNs) 
are one of the candidates of non-thermal emission sources \citepe{2018ApJ...859..164B,2019MNRAS.490..978P}. \cite{2018ApJ...859..164B} find that AGN models can reproduce observed {\HeII\W1640},
while that other emission lines, such as {\CIV\W\W1548,1551} and {\CIIIsemi\W1909} cannot be reproduced simultaneously. However, their AGN models does not consider metal-poor ($<0.25$ $Z_\odot$) AGNs. Shocks are another candidates for the ionizing sources of {\HeII} \citepe{2005ApJ...621..269T,2021ApJ...918...54I,2018ApJ...859..164B,2019MNRAS.490..978P}. \cite{2018ApJ...859..164B} also examine contributions from radiative shocks using radiative shock models of \cite{2008ApJS..178...20A}. The calculations with the radiative shock models also suggest that radiative shocks cannot explain the strong {\HeII\W1640} emission for a given intensity of other emission lines such as {\CIV\W\W1548,1551} and {\CIIIsemi\W1909} (see also \citealt{2021MNRAS.tmp.2563I}).
\par
Alternatively, other hard radiation sources are also proposed. \cite{2019A&A...622L..10S} show that X-ray luminosities have a positive correlation with line ratios {\HeII\W4686}/{\Hb} with X-ray binary population models, indicating the contributions of the {\HeII} ionizing photons from high mass X-ray binaries (HMXBs). \cite{2017AA...602A..45L} demonstrate that \Add{HMXBs} can be the sources of {\HeII} ionizing photons in \Add{well-studied EMPG} I Zw 18 through detailed modeling of radiation fields and interstellar mediums. \Add{\cite{2021ApJ...908L..54K} argue that the ultra-luminous X-ray source (ULX) in I Zw 18 cannot fully account for all of {\HeII} ionizing photons. However, new integral field spectroscopic observations of I Zw 18 suggest that the ULX still remains the possibility \citep[see][]{2021ApJ...911L..17R}.} \Add{Moreover, while} \cite{2020MNRAS.494..941S} argue that HMXBs cannot be the main contributors of {\HeII} \Add{in EMPGs} through the photoionization models representing HXMB spectra by simple multi-color disk models, \cite{2021arXiv210812438S} examine the photoionization models using several detailed spectrum models for HMXBs/\Add{ULXs} and find that some of the models can reproduce {\HeII\W4686}/{\Hb} and other line ratios such as {\OIII\W\W4959,5007}/{\Hb} and {\OIII\W\W4959,5007}/{\OII\W3727}. The studies of \cite{2019A&A...622L..10S}, \cite{2021arXiv210812438S}, and \cite{2017AA...602A..45L} claim that HMXB/ULX can explain strong {\HeII} emission lines (cf. \citealt{2020saxena} and \citealt{2020MNRAS.494..941S}), but whether a simple physical model can explain all the emission lines from a galaxy in a self-consistent manner remains uncertain. There are also attempts to explain emission lines from EMPGs by non-uniform inter-stellar medium (ISM) \citepe{2021arXiv210512765B,2021MNRAS.tmp.2563I,2020A&A...644A..21R}. \Add{For example, \cite{2020A&A...644A..21R} show that ordinary stellar population synthesis models can reproduce high-ionization emission lines {\OIII\W\W4959,5007} simultaneously with low-ionization emission lines {\OII\W3727} by introducing so-called ``two-zone" models.}\par
Variations of models presented by these previous studies raise difficulties in reproducing strong {\HeII} emission line fluxes for given low-ionization line fluxes. Instead, we aim to determine the self-consistent general spectrum shape by a parametric method. By obtaining the general spectrum shapes first, we can infer the possibility of countless ionization sources without running costly calculations using photoionization code each time. Moreover, we expect to uncover the spectral features that were not previously mentioned.\par
In this work, we search for the self-consistent spectrum shapes that can reproduce observed emission line fluxes of three EMPGs. Similar work such as \cite{2021arXiv210906725O} (hereafter, O21) uses a blackbody and BPASS models to search for the spectrum shape. Another similar study uses combinations of BPASS models and ULX models as ionizing spectra \citep{2021arXiv210812438S}. With the minimum assumptions, we construct generalized spectral models that cover spectrum shapes examined in \citetalias{2021arXiv210906725O} and \cite{2021arXiv210812438S}. We use a generalized spectrum as an ionizing spectrum in a uniform ISM to compute emission line fluxes using photoionization code {\sc cloudy}. We then search for the self-consistent spectrum shapes using Markov Chain Monte Carlo (MCMC) method.\par
Our sample and methodology are described in Section \ref{sec:methods}. We present our results for the general spectrum shapes and reproduction of emission lines in Section \ref{sec:results}. In Section \ref{sec:discuss}, we evaluate several candidates for the hard radiation sources using the general spectrum shape. We also examine the contributions of the hard radiation sources to cosmic reionization. In Section \ref{sec:conc}, we summarize our main results. Throughout this paper, magnitudes are in the AB system, and we assume a standard $\Lambda$ CDM cosmology with parameters of ($\Omega_{\rm m}$, $\Omega_{\rm \Lambda}$, $H_{0}$) = (0.3, 0.7, 70 \kms ${\rm Mpc}^{-1}$) and the solar metallicity scale of \cite{2009ARA&A..47..481A}, where \metal=8.69.\par
\input{galaxy.tex}
\section{Sample and Methods} \label{sec:methods}

\subsection{Sample}

\begin{figure}[t]
\begin{center}
    \includegraphics[width=\linewidth,trim=5mm 3mm 0mm 0mm,clip]{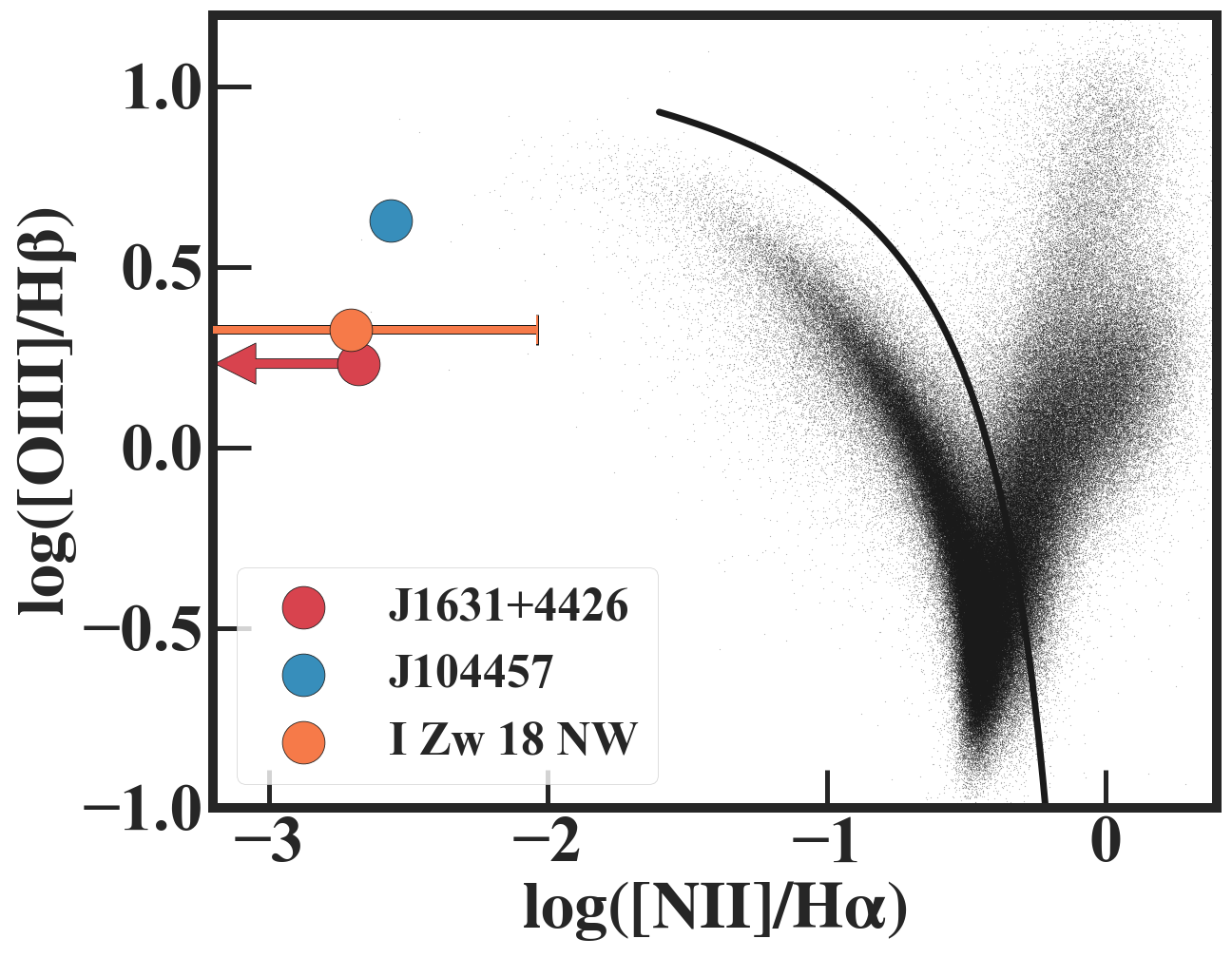}
\caption{Our three galaxies on the BPT diagram. The red, blue, and orange dots represent the emission line ratios of {\EMPGa}, {\EMPGb}, and {\EMPGc}, respectively. The bars represent errors of the emission line ratios. The red arrow indicates an upper limit with $1\sigma$ level for {\NII\W6583}/{\Ha} of {\EMPGa}. The black dots represent $z\sim0$ SFGs and AGNs derived from the emission line catalog of SDSS DR7. The solid curve depicts the maximum photoionization models that can be reproduced solely by stellar radiation \citep{2003MNRAS.346.1055K}. The region left (right) to the solid curve is the SFG (AGN and composite) region.}
\label{BPT}
\end{center}
\end{figure}

We investigate general ionizing spectrum shapes for a sample of three galaxies, {\EMPGa}, {\EMPGb}, and {\EMPGc}, whose properties are summarized in Table \ref{table:gal}. 
All three galaxies are extremely metal-poor ($Z<10\%~Z_\odot$), nearby ($z\sim0$), and low-mass ($\lesssim 10^6 M_\odot$) galaxies. J1631+4426 is the most metal-poor galaxy detected, with $Z=1.6\%~Z_\odot$ \citep[][hereafter, K21]{2021ApJ...913...22K}. \Add{Thus, {\EMPGa} is a good test-bed for investigating the nature of EMPGs.} {\EMPGb} is a metal-poor galaxy, whose optical spectrum has been studied previously and shown that BPASS models alone cannot explain high-energy ionization lines \citep[][hereafter, B21]{2021arXiv210512765B}. \citet{2021arXiv210906725O} have also studied \Add{{\EMPGb} and have shown that the combination of a blackbody and BPASS model can explain multi-level emission lines of {\EMPGb}. Therefore, we can compare our results to those of \citet{2021arXiv210906725O}.} I Zw 18 NW is one of the most well-known EMPG that has optical and X-ray data \citep[][hereafter, TI05]{2005ApJS..161..240T}. \Add{This allows us to check the consistency between the ionizing spectrum shapes beyond 54.4 eV and X-ray luminosity data for {\EMPGc}.} All three galaxies in our sample have optical spectroscopic data available. In Figure \ref{BPT}, we present the locations of all three galaxies on a diagram of {\NII\W6583}/{\Ha} and {\OIII\W5007}/{\Hb} emission line ratios, so-called the Baldwin-Philips-Terlevich diagram \citep[BPT diagram;][]{1981PASP...93....5B}. The black dots represent the emission line ratios of $z\sim0$ star forming galaxies (SFGs) and AGNs from the Sloan Digital Sky Survey (SDSS) Data Release 7 (DR7) \citep{2009ApJS..182..543A}. {\NII\W6583}/{\Ha} and {\OIII\W5007}/{\Hb} emission line ratios for all three galaxies can be explained by only stellar radiation \citep{2003MNRAS.346.1055K}. However, because \cite{2013ApJ...774..100K} suggest that metal-poor gas heated by AGN can produce emission line ratio values that fall on the SFG region, we cannot conclude the contributions from AGNs from this classification.
\par
In this work, we use observed optical emission lines from hydrogen, helium, oxygen, and sulfur ions to determine the best-fit model parameters. The optical emission lines and their flux ratios are listed in Table \ref{table:flux1}.
We avoid using {\Ha} for the fitting of J1631+4426 because the reported {\Ha} flux is not consistent with the Balmer decrements. We also avoid using {\Ha} for the fitting of I Zw 18 NW because the value of {\Ha} flux has been obtained using a different spectrometer from the one used to obtain the fluxes of the other lines. We use line ratios between sulfur doublets {\SII}\W6316/\W6731 instead of line ratios between sulfur and hydrogen to minimize effects from the uncertainty of sulfur's abundance ratio to oxygen. Another reason to use {\SII}\W6316/\W6731 is that this line ratio is sensitive to the gas density in {\HII} regions \citep{2006agna.book.....O}. We avoid including emission line fluxes from other elements to minimize systematics of abundance ratios. We note that {\HeII}\W4686 lines of {\EMPGa} and {\EMPGb} are narrow and have no broad-line features \citep{2021ApJ...913...22K,2021arXiv210512765B}. There is a report of a Wolf-Rayet bump centered at 4645 {\AA} for {\EMPGc}, but no underlying broad component is detected for {\HeII}\W4686 \citep{1997A&A...326L..17L}. Therefore, we treat {\HeII}\W4686 lines of our three galaxies as the nebular origin. All the line ratios in Table \ref{table:flux1} use the dust-corrected fluxes.\par
%
\input{flux1.tex}

\subsection{Photoionization Models}
We use the spectral synthetic code {\sc cloudy} \citep[version 17.02;][]{2017RMxAA..53..385F} to calculate the emission line fluxes from a gas cloud ionized by a central ionizing source. With  {\sc cloudy}, we simulate physical conditions within a gas cloud to predict the emission line fluxes. To obtain the emission line 
fluxes, we use photoionization models with free parameters of ionizing spectra and nebulae. We describe these free parameters as well as fixed parameters below.
\subsubsection{Geometry and Density}
In our photoionization model, we assume a spherical shell of the gas cloud surrounding the central ionizing source, in accordance with the default setting of {\sc cloudy.\footnote{More descriptions in the {\sc cloudy} documentation Hazy 1}} We fix the inner radius \Add{(outer radius)} of the gas cloud $R_{\rm{in}}$ \Add{($R_{\rm{out}}$)} at the default value. \Add{The default value of $R_{\rm{out}}$ is at $R_{\rm{out}}=10^{30}~\rm{cm}$, and} this large \Add{$R_{\rm{out}}$ value} produces an effectively plane-parallel geometry. For simplicity, we assume the gas cloud with the constant hydrogen density $n_{\rm{H}}$. To cover a range of typical $n_{\rm H}$ values for the hydrogen density of star-forming galaxies, we apply the hydrogen density ranging in ${10^{0.5}}\leq {n_{\rm{H}}} \leq {10^{5}~{\rm cm}^{-3}}$.\par
\subsubsection{Ionizing Spectra}
Previous studies have examined ionizing spectra of a blackbody, stellar, and hard radiation sources (e.g., ULXs, and combinations of these sources \citepe{2021arXiv210906725O,2021arXiv210812438S,2021MNRAS.tmp.2881F}. However, for all the soft and hard radiation models, whether a single self-consistent model can explain emission line fluxes from a single galaxy remains unclear. To solve this, we use a generalized ionizing spectrum composed of a blackbody and power-law radiation. The flux of this spectrum at the frequency $\nu$ is given as
\begin{equation}
    F_\nu = B(\nu,~T_{\rm BB})+C_{\rm mix}P(\nu,~\alpha_{\rm X}).
\end{equation}
$B(\nu,~T_{\rm BB})$ and $P(\nu,~\alpha_{\rm X})$ are given as
\begin{eqnarray}
B(\nu,~T_{\rm BB}) & = & \frac{h\nu^3}{\exp(h\nu/kT_{\rm{BB}})-1} \\
P(\nu,~\alpha_{\rm X}) & = & \nu^{\alpha_{\rm X}}\exp(h\nu/E_{\rm{lc}})\exp(-h\nu/E_{\rm{hc}}), 
\end{eqnarray}
where $T_{\rm{BB}}$, $C_{\rm mix}$, $\alpha_{\rm X}$, $h$ and $k$ are the blackbody temperature, the mixing parameter, the power-law index, the Planck constant, and the Boltzmann constant, respectively. For the power-law component, we define the lower (higher) cut energy $E_{\rm{lc}}$ ($E_{\rm{hc}}$). 
We set $E_{\rm{lc}}=0.1~\rm{Ryd}$ ($E_{\rm{hc}}=10000~\rm{Ryd}$) to avoid strong free-free (pair-creation and Compton) heating. We limit the blackbody temperature and power-law index in a range of $4\leq \log T_{\rm BB}~\leq6$ and $-3\leq \alpha_{\rm X} \leq 1$, respectively. For $C_{\rm mix}$, we introduce the alternative parameter $a_{\rm mix}$ defined as

\begin{equation}
    a_{\rm mix} \equiv C_{\rm mix}\frac{P(\nu=1~{\rm Ryd}/h, \alpha_{\rm X})}{B(\nu=1~{\rm Ryd}/h, T_{\rm BB})}.
\label{mixing}
\end{equation}
In our model calculations, we specify a value for $a_{\rm mix}$ instead of a value for $C_{\rm mix}$ to match the specification of {\sc cloudy}. We allow $a_{\rm mix}$ to vary in a range of $-3 \leq \log a_{\rm mix} \leq 3$. Our generalized spectral models cover various spectrum shapes. For example, ($T_{\rm BB}$, $\alpha_{\rm X}$, $a_{\rm mix}$)=($10^5~{\rm K}$, -1.8, -0.1) can mimic the FUV spectrum shape of an AGN model generated by {\sc cloudy} ``AGN" continuum command with the default parameters.\par
\subsubsection{Chemical Abundance}

We use solar abundance ratios of \cite{2010Ap&SS.328..179G} (i.e., GASS10) as an initial condition of our {\sc cloudy} model calculations. 
We scale all the abundances linearly with the oxygen abundance (hereafter referred to as metallicity) $Z$, except for those of helium, carbon, and nitrogen. We assume that the helium and carbon abundances follow the non-linear relationships that are given by \cite{2006ApJS..167..177D}. For the scaling of nitrogen with metallicity, we use the formula given by \cite{2012MNRAS.426.2630L}. We adopt a metallicity range of $-3 \leq \log~Z/Z_\odot \leq 0$. Note that we do not use emission lines from carbon and nitrogen, so the uncertainty for these ions' abundances will only have minor effects {on} our results.\par
\subsubsection{Ionizing Spectrum Intensity}
The intensity of the ionizing spectrum is specified with the ionization parameter $U$. The ionization parameter is defined as 
\begin{equation}
    U\equiv \frac{Q({\rm H}^{0})}{4\pi {R_{\rm S}}^2 n_{\rm H} c},
    \label{ion_para}
\end{equation}
where $Q({\rm H}^{0})$, $R_{\rm S}$, and $c$ are the intensity of ionizing photons above the Lyman limit ($\leq 912~${\AA}), the Str\"omgren radius, and the speed of light, respectively. Because $R_{\rm S}$ is nearly equal to $R_{\rm in}$ in the geometry of our model, we approximate $U$ by substituting $R_{\rm S}=R_{\rm in}$. We impose a flat prior for $U$ in a range of $-5 \leq \log~U \leq -0.5$.\par
\subsubsection{Other Model Specifications}
Our models are truncated at a neutral hydrogen column density $N_{\rm{HI}}$ at $N_{\rm{HI}}=10^{21}~\rm{cm}^{-2}$.
We normalize the output emission line fluxes relative to the model $\rm{H\beta}$ flux for convenience of calculation.\par
\subsection{Markov Chain Monte Carlo methods}
To estimate the best-fit parameters, we combine our {\sc cloudy} photoionization models with Markov Chain Monte Carlo (MCMC) methods. To run the MCMC methods, we use {\tt\string emcee} \citep{2013RMxAA..49..137F}, a Python implementation of an affine invariant MCMC sampling algorithm. We use the log-likelihood Gaussian function in our MCMC methods. 
Our log-likelihood function $\ln\mathcal{L}$ is:
\begin{equation}
    \ln\mathcal{L} = -\frac{1}{2}\sum_{\lambda \in \Lambda}\left[ \left(
    \frac{F_{\lambda,\rm{obs}}-F_{\lambda,\rm{mod}}}{\sigma_{\lambda,\rm{obs}}}
    \right)^2+\ln(2\pi \sigma_{\lambda,\rm{obs}}^2)\right],
    \label{logL}
\end{equation}
where $\Lambda$, $F_{\lambda,\rm{obs}}$, $F_{\lambda,\rm{mod}}$, and $\sigma_{\lambda,\rm{obs}}$ are the set of emission lines used, the observed emission line flux at wavelength $\lambda$, the model emission line flux at wavelength $\lambda$, and the observed error of the emission line at wavelength $\lambda$, respectively. $F_{\lambda,\rm{mod}}$ in equation \ref{logL} is given by
\begin{equation}
    F_{\lambda,\rm{mod}}=\mathit{N}_{\rm{H\beta}}\frac{F_{\lambda,\rm{ cloudy}}}{F_{\rm{H\beta},\rm{ cloudy}}},
\end{equation}
where $F_{\lambda, \rm{cloudy}}$ and $\mathit{N}_{\rm{H\beta}}$ are the output of {\sc cloudy} for the emission line flux at the wavelength $\lambda$ and the normalization factor for an $\rm{H\beta}$ emission line, respectively. \Add{$F_{\lambda,\rm{ cloudy}}$ values are normalized at $F_{\rm{H\beta},\rm{ cloudy}}=1$ by default in {\sc cloudy}.} We introduce $\mathit{N}_{\rm{H\beta}}$ as a new free parameter to scale the model fluxes to the observed fluxes. We allow $\mathit{N}_{\rm{H\beta}}$ \Add{(i.e., $F_{{\rm H\beta},\rm{mod}}$ by definition)} to vary with in a range of $3\sigma_{\rm{H\beta},\rm{obs}}$ from $F_{\rm{H\beta},\rm{obs}}$. \Add{Here, $F_{\rm{\lambda},\rm{obs}}$ values are normalized at $F_{\rm{H\beta},\rm{obs}}=100$ as shown in Table \ref{table:flux1}.} We have a total of seven free parameters in our photoionization models. As a prior probability distributions for each model parameters, we adopt a simple uniform distributions in the range permitted for each variable. We summarize the prior distributions for all the free parameters in Table \ref{table:prior}. \par
We initialize the positions of the parameters by randomly selecting values from the prior distributions. We use 40 walkers and let each chain run for 1000 steps. We define the ``best-fit" parameter set as a parameter set with the maximum likelihood value in the sampled parameter sets. The uncertainty is defined by the range between the minimum and maximum values of the parameter sets that satisfy a condition

\begin{equation}
\begin{aligned}
\ln\mathcal{L} & \geq \ln\mathcal{L}_{3\sigma}\\ 
& \equiv -\frac{1}{2}\sum_{\lambda \in \Lambda}\left[ \left(
\frac{3\sigma_{\lambda,\rm{obs}}}{\sigma_{\lambda,\rm{obs}}}
\right)^2+\ln(2\pi \sigma_{\lambda,\rm{obs}}^2)\right]. \\
\label{uncert}
\end{aligned}
\end{equation}
Note that this uncertainty is only a rough standard and the actual uncertainty could be more strictly given. \par
\input{prior.tex}
\section{Results}\label{sec:results}
\subsection{Posterior Probability Distribution Functions}
\begin{figure*}[hbt!]
\begin{center}
\includegraphics[width = \linewidth,clip]{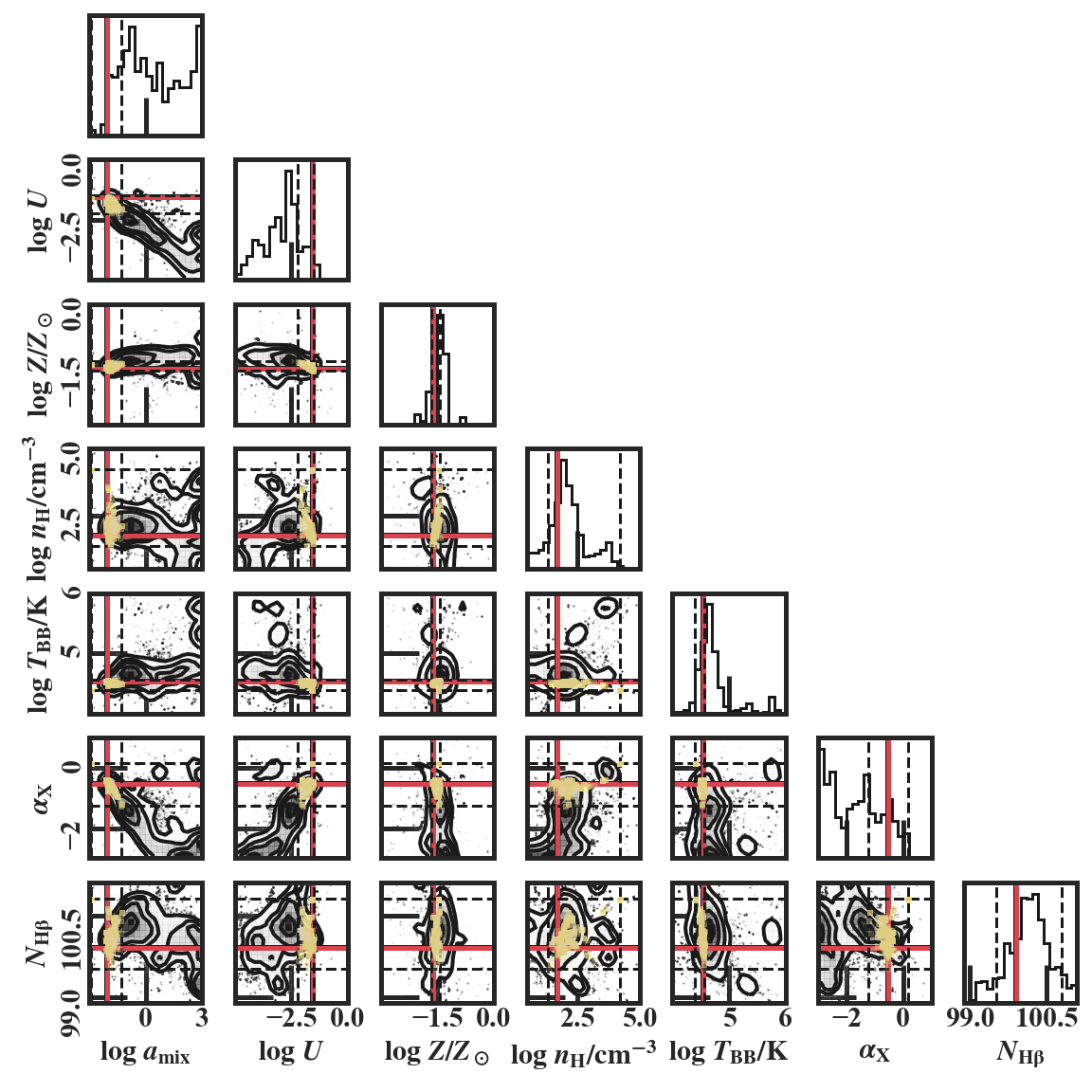}
\caption{Posterior PDF-triangles for {\EMPGa}. One-dimensional PDFs are shown along the diagonals, while the two-dimensional PDFs are shown on the off-diagonals. The darker regions on the two-dimensional PDFs indicate the higher density of the parameter sets. The red lines (black dashed lines) indicate the values of the best-fit parameters (uncertainty ranges). The yellow dots are the sampled parameter sets within the uncertainty ranges.}
\label{PDFa}
\end{center}
\end{figure*}

\begin{figure*}[hbt!]
\begin{center}
\includegraphics[width = \linewidth,clip]{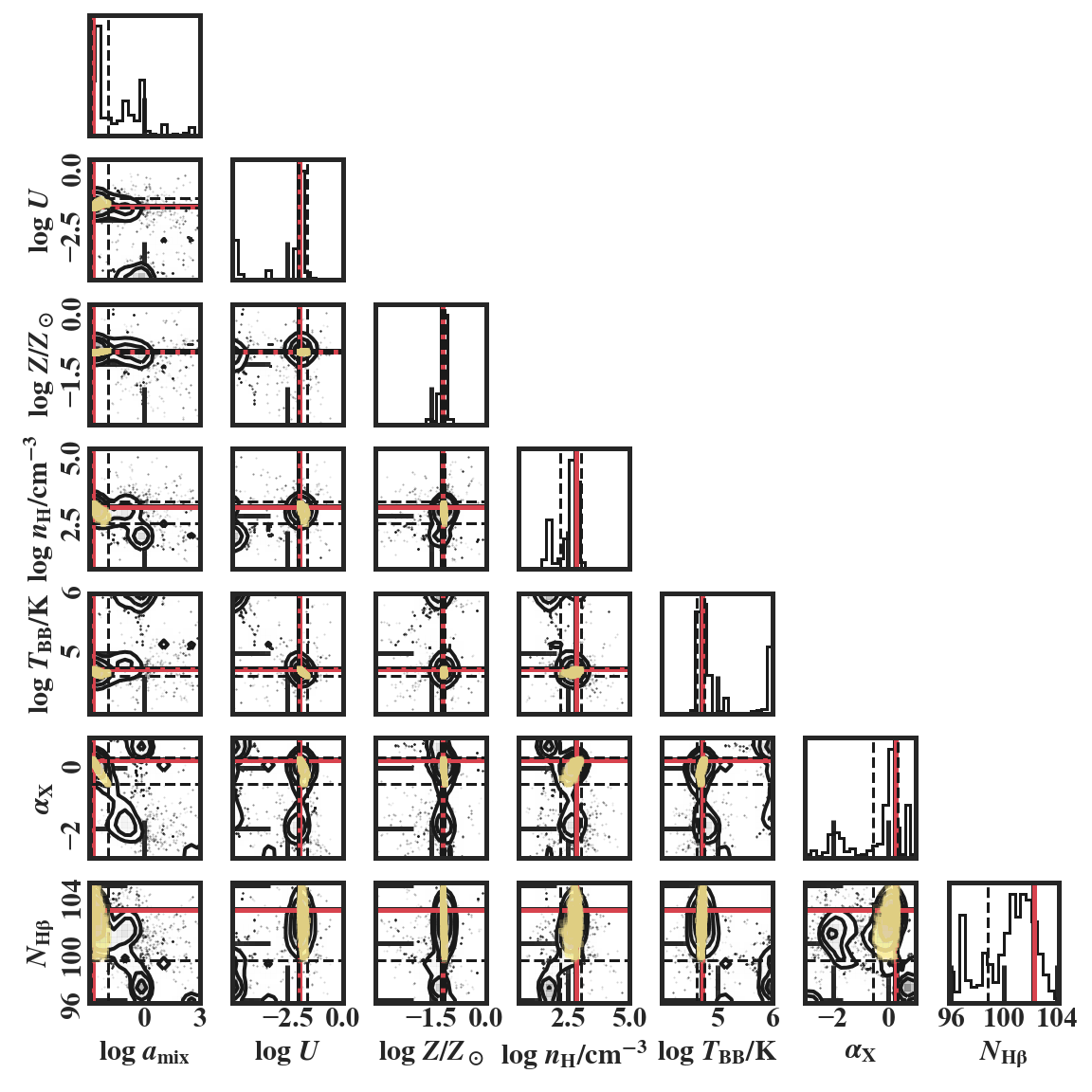}
\caption{Posterior PDF-triangles for {\EMPGb}. One-dimensional PDFs are shown along the diagonals, while the two-dimensional PDFs are shown on the off-diagonals. The darker regions on the two-dimensional PDFs indicate the higher density of the parameter sets. The red lines (black dashed lines) indicate the values of the best-fit parameters (uncertainty ranges). The yellow dots are the sampled parameter sets within the uncertainty ranges.}
\label{PDFb}
\end{center}
\end{figure*}

\begin{figure*}[hbt!]
\begin{center}
\includegraphics[width = \linewidth,clip]{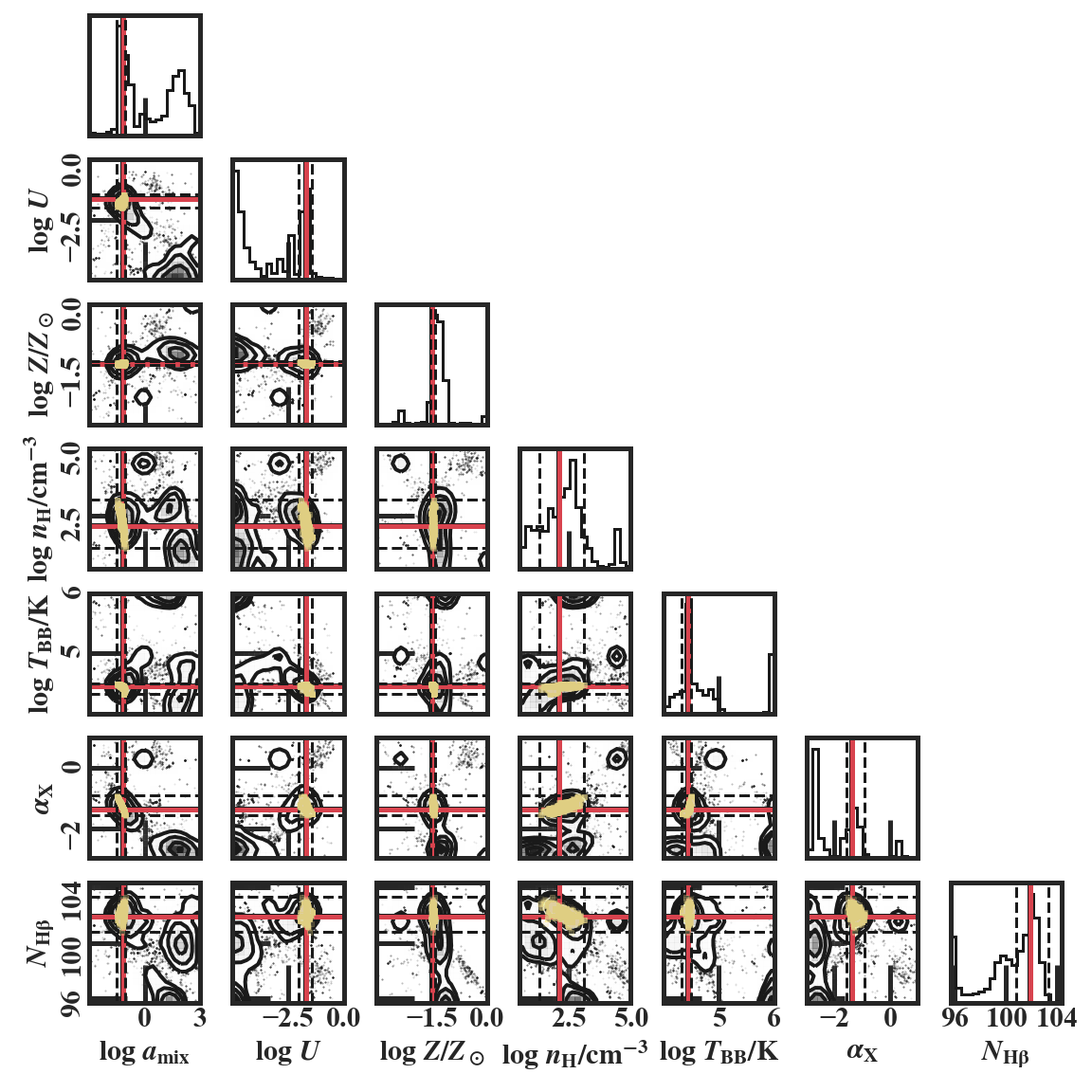}
\caption{Posterior PDF-triangles for {\EMPGc}. One-dimensional PDFs are shown along the diagonals, while the two-dimensional PDFs are shown on the off-diagonals. The darker regions on the two-dimensional PDFs indicate the higher density of the parameter sets. The red lines (black dashed lines) indicate the values of the best-fit parameters (uncertainty ranges). The yellow dots are the sampled parameter sets within the uncertainty ranges.}
\label{PDFc}
\end{center}
\end{figure*}

We obtain the best-fit parameters for all three galaxies in our sample using the method described in Section \ref{sec:methods}. We present the posterior probability distribution functions (PDFs)-triangles for {\EMPGa}, {\EMPGb}, and {\EMPGc} in Figure \Add{\ref{PDFa}, \ref{PDFb}, and \ref{PDFc}}, respectively.\par

One-dimensional and two-dimensional PDFs are plotted along the diagonal and on the off-diagonal, respectively. The grey scale in each of the two-dimensional PDFs represents the probability density of the sampled parameters. The darker region on the two-dimensional PDFs indicates a higher density of the sampled parameter sets. The red lines (the black dashed lines) indicate the values of the best-fit parameters (uncertainty ranges). The yellow dots are the sampled parameter sets within the uncertainty ranges. We summarize the best-fit parameters and their uncertainties in Table \ref{table:best_params}.\par

In Figure \Add{\ref{PDFa}, \ref{PDFb}, and \ref{PDFc}}, the prior ranges are wide enough to reveal the overall shapes of the PDFs that are not distorted by the prior boundaries except for that of $\log a_{\rm mix}$ of {\EMPGb}. However, the uncertainty ranges for $\log a_{\rm mix}$ of {\EMPGb} are within its prior range. The PDF for some of the parameters has multi-modal distribution. In contrast, parameter sets within the uncertainty ranges (yellow dots) are concentrated around the best-fit parameters. This suggests that peaks other than those of the best-fit parameters are just local maxima of equation \ref{logL}. The best-fit values for metallicity of our three galaxies are comparable (within 0.2 dex) with the literature values of metallicity.\par
\subsection{Reproduction of Emission Lines}
\begin{figure}[t]
\includegraphics[width = \linewidth,clip]{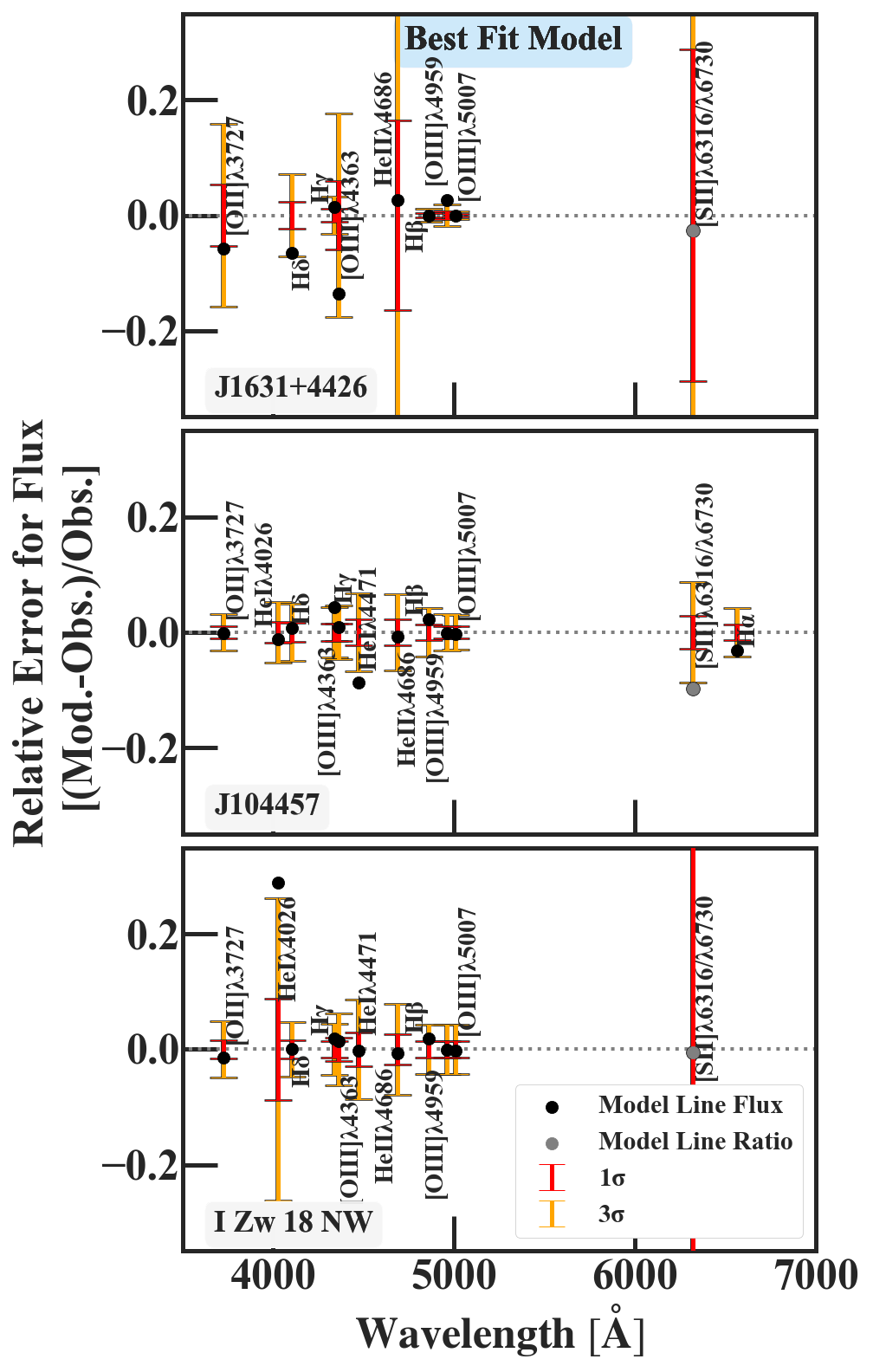}
\caption{Emission line fluxes and flux ratios reproduced from the best-fit parameters. We show results for {\EMPGa} (top), {\EMPGb} (middle), and {\EMPGc} (bottom). The relative errors for fluxes (black dots) are differences between the value produced from the best-fit parameters and the observed values, normalized by the observed values. The red (yellow) bars represent the $1\sigma$ ($3\sigma$) errors of the observed values.}
\label{el}
\end{figure}

\input{best_params.tex}
We compare the observed emission line fluxes with the model emission line fluxes reproduced with the best-fit parameters. In Figure \ref{el}, we present normalized differences between the observed and the model fluxes for emission lines in the black dots. We also plot the $1\sigma_{\lambda,\rm{obs}}$ and $3\sigma_{\lambda,\rm{obs}}$ ranges for the emission lines in the red and yellow bars, respectively. From Figure \ref{el}, we find that the best-fit models simultaneously reproduce all the emission lines within around $3\sigma_{\lambda,\rm{obs}}$ for all of our three galaxies. \Add{Interestingly, {\OIII}\W4363 is the most deviant line for {\EMPGa}. This deviance is probably because our current model for the MCMC method could be weighting too much on reproducing {\OIII}\W\W4959,5007. The 1$\sigma$ values for {\OIII}\W\W4959,5007 are below 1\% of the flux values, while 1$\sigma$ values for {\OIII}\W4363 is about 5\% of the flux value.}\par

\subsection{Ionizing Spectrum Shapes}
\begin{figure*}[p]
\begin{center}
    \includegraphics[width=\textwidth,trim=5mm 3mm 0mm 0mm,clip]{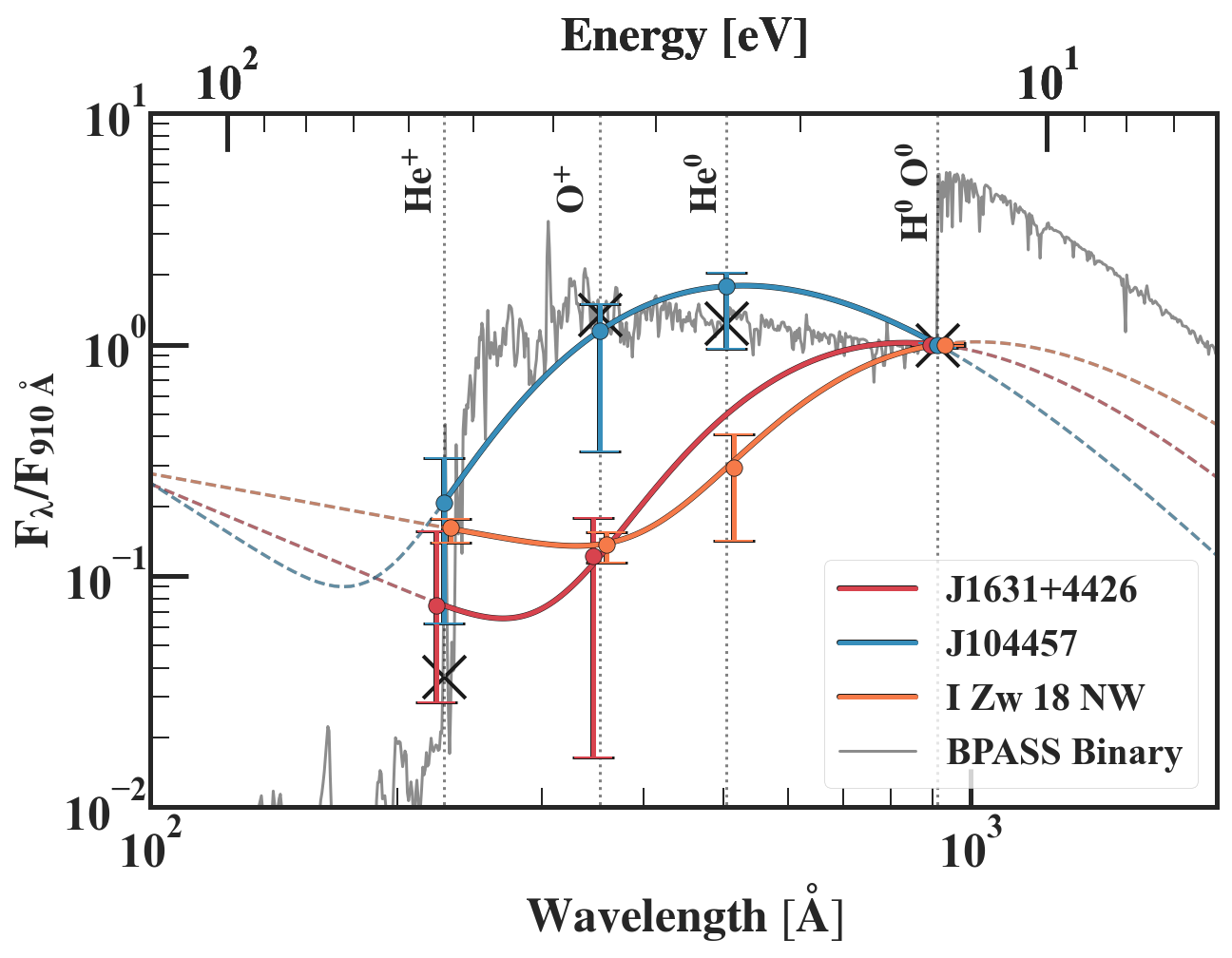}
\caption{Ionizing spectrum shapes reproduced for {\EMPGa} (red line), {\EMPGb} (blue line), and {\EMPGc} (orange line) with our MCMC methods. The grey line is an example ionizing spectrum from the BPASS binary model (${\rm Age}=10~{\rm Myr},~ Z_*=0.05Z_{\odot}$). The vertical dotted lines mark the ionization energies of ${\rm H^{0}}$, ${\rm O^{0}}$, ${\rm He^{0}}$, ${\rm O^{+}}$, and ${\rm He^{+}}$. The colored dot (bar) represents the value of flux (the range of uncertainty) at each ionization energy for the best-fit ionizing spectra. The black cross represents the value of flux at each ionization energy for the BPASS binary model. All ionizing spectra are normalized at $910~${\AA}}
\label{seds}
\end{center}
\end{figure*}

In Figure \ref{seds}, we depict the general ionizing spectrum shapes reproduced from the best-fit parameters for our three galaxies . Hereafter, we call these ionizing spectra as the ``best-fit" ionizing spectra. For comparison, we overplot the ionizing spectrum of the BPASS binary model at the stellar age 10 Myr and the stellar metallicity $Z_*$ at $0.05~Z_{\odot}$. We use BPASS v2.2.1 \citep{2017PASA...34...58E,2018MNRAS.479...75S} with a Kroupa initial mass function (IMF) and a high mass cut-off at $300~M_{\odot}$. All ionizing spectra are normalized at 910 {\AA}. We also plot uncertainties for the spectral fluxes at ionization energies of ions used in our parameter search. The uncertainty bars are given by the maximum and the minimum spectral flux values reproduced with the sampled parameter sets satisfying equation \ref{uncert}. We note again that these uncertainties are only rough standards and the actual uncertainties could be be more strictly given.\par

Between the ionization energies of ${\rm H^0}$ $(13.6~{\rm eV})$ and ${\rm He^+}$ $(54.4~{\rm eV})$, we find that the best-fit ionizing spectra have two types of general spectrum shapes. The best-fit ionizing spectrum for {\EMPGb} has a convex upward spectrum shape like the one for BPASS binary models, except that most of the {\HeII} ionizing photons are absorbed by the stellar atmosphere in the BPASS binary models. On the other hand, the best-fit ionizing spectra for {\EMPGa} and {\EMPGc} have convex downward spectrum shapes that BPASS binary models cannot reproduce. These convex downward spectrum shapes have dominant contributions from the power-law radiation around the ionization energy of ${\rm He}^{+}~(54.4~{\rm eV})$. These two types of general spectrum shapes indicate a diversity of the ionizing spectrum shapes for EMPGs.\par

\subsection{Comparison with \Add{Other Models}}
\begin{figure*}[htbp]
\gridline{\fig{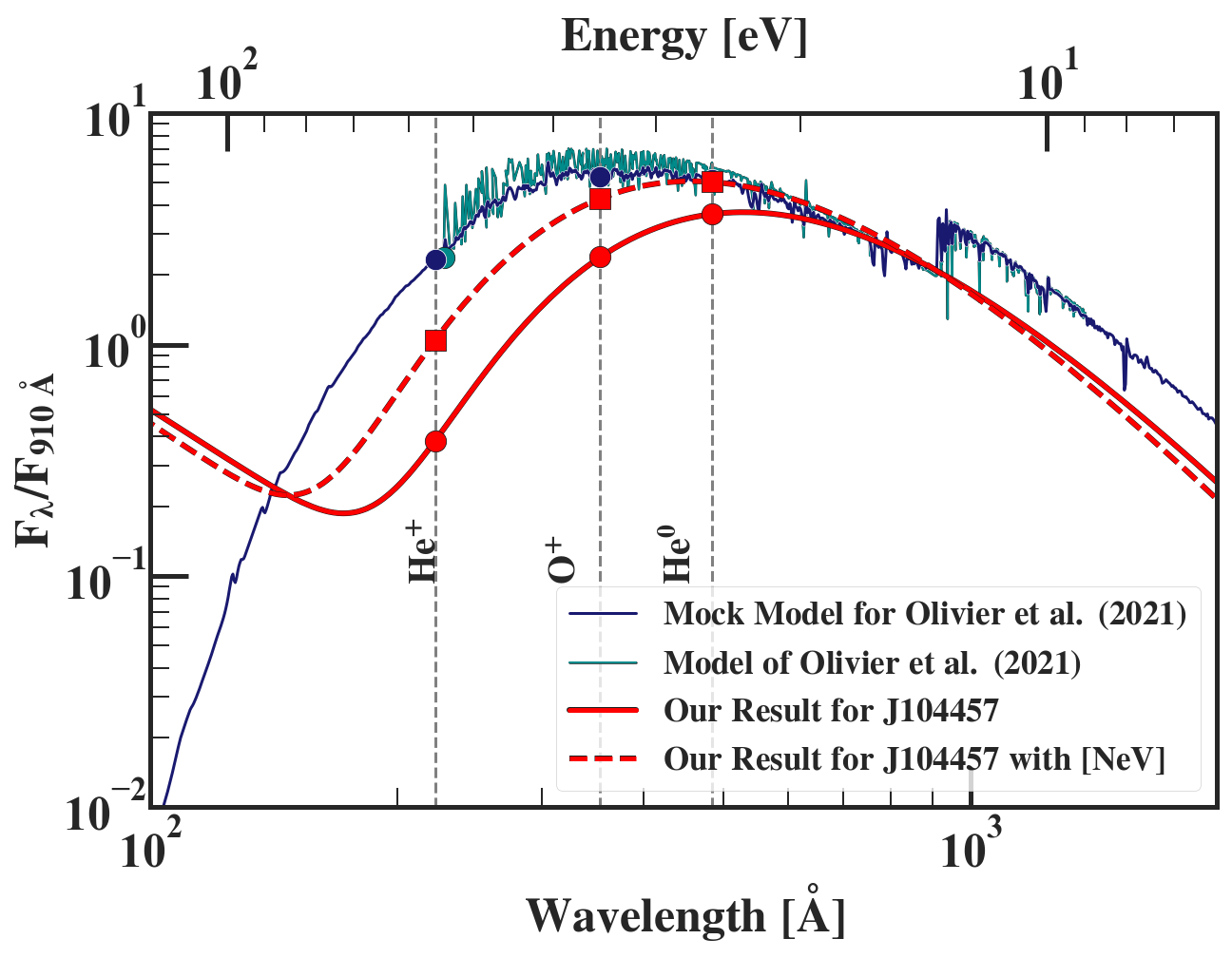}{0.45\textwidth}{(a)}
\fig{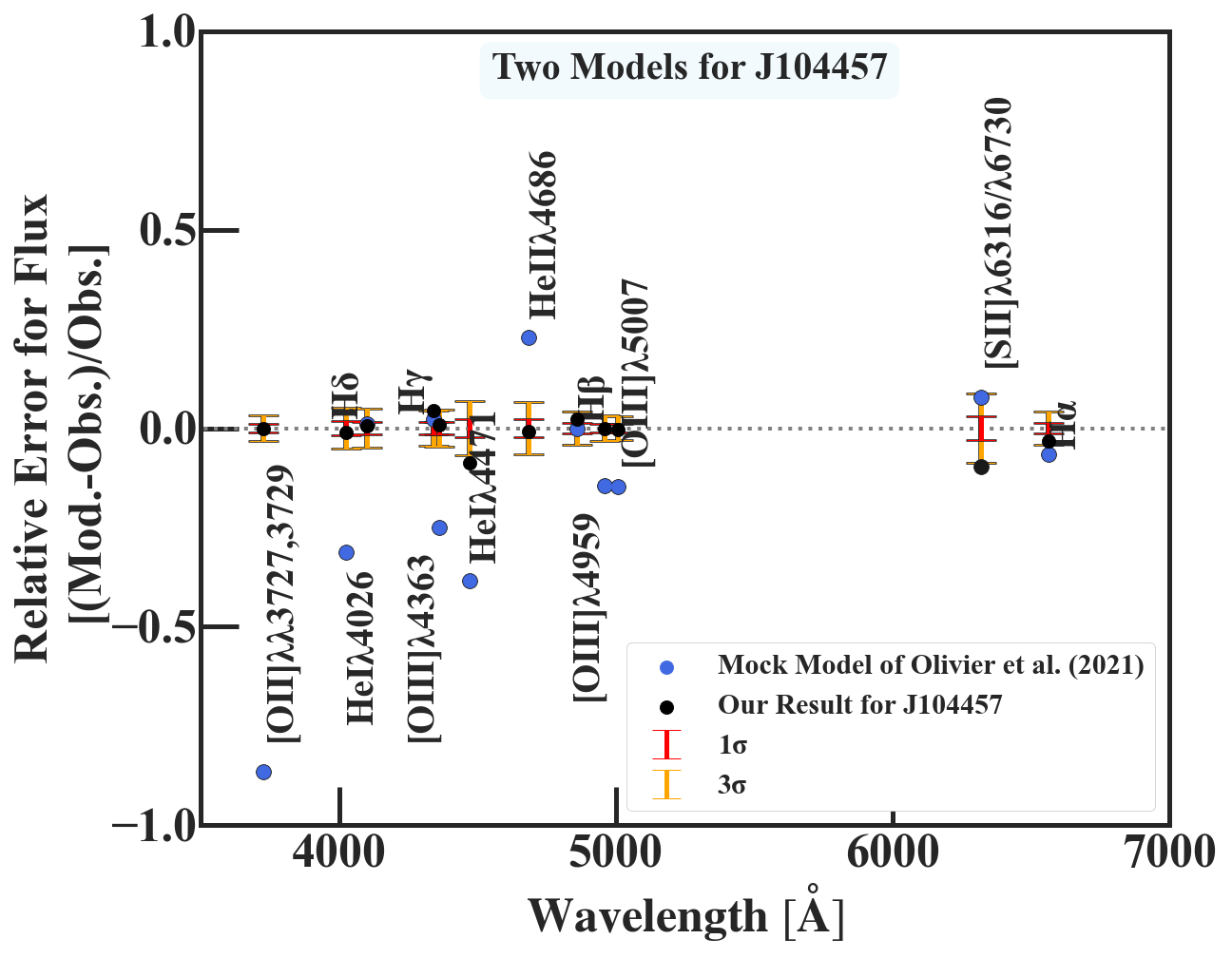}{0.45\textwidth}{(b)} 
          }
\caption{Comparison between our result for {\EMPGb} and \Add{other models}. (a) The blue, green, \Add{red solid, and red dashed} lines are the spectrum of the mock model for \citetalias{2021arXiv210906725O}, the model in \citetalias{2021arXiv210906725O}, our result for {\EMPGb}, \Add{and our result when we include {\NeV}\W3426/{\NeIII}\W3869}, respectively. Only the spectrum below 54.4 eV is shown for the model in \citetalias{2021arXiv210906725O}. The other symbols are the same as those in Figure \ref{seds}. (b) Emission line fluxes and emission line ratios from two models. The blue (back) dots are the emission line fluxes and emission line ratio from the mock model of \citetalias{2021arXiv210906725O} (our result for {\EMPGb}). The other symbols are the same as those in Figure \ref{el}.}
\label{olivier}
\end{figure*}

\citetalias{2021arXiv210906725O} claim that the combination of a BPASS binary model and blackbody radiation can reproduce emission lines from {\EMPGb} in a self-consistent manner. We compare our self-consistent model for {\EMPGb} with the model for {\EMPGb} presented in \citetalias{2021arXiv210906725O}. We construct a mock model for the model presented \citetalias{2021arXiv210906725O} using the model parameters listed in the bottom column of Table 5 in the same paper. We use the same BPASS version and initial mass function as \citetalias{2021arXiv210906725O} for the mock model. In Figure \ref{olivier} (a), we present a spectrum of the mock model in \Add{the} blue \Add{line}, the spectrum from \citetalias{2021arXiv210906725O} in \Add{the} green \Add{line}, and our result for {\EMPGb} in \Add{the} red \Add{solid line}. Only the spectrum for below 54.4 eV is shown for the model in \citetalias{2021arXiv210906725O}. We find that the spectrum in \citetalias{2021arXiv210906725O} has relatively higher fluxes around 54.4 eV than the spectrum of our result.

In Figure \ref{olivier} (b), we show emission line fluxes and emission line ratios for the mock model (our model) in blue (black) dots. The mock model of \citetalias{2021arXiv210906725O} reproduces the observed {\HeII\W4686} fairly well. However, other emission lines, especially the low-ionization emission line fluxes such as {\OII}\W\W3727,3729, {\HeI\W}4026, and {\HeI\W}4471, are systematically underproduced, unlike our result. One possible explanation for this underproduction by the mock model is that the general spectrum shape for the mock model has a relatively lower ratio of low-energy ($13.6-24.6$ eV) fluxes to high-energy ($35.1-54.4$ eV) fluxes than that of our result.\par
\Add{Next, we consider how taking {\NeV} emission lines into consideration affects our result. {\NeV} emission lines are detected in some of the {\HeII} emitting EMPGs including {\EMPGb} \citepe{2004A&A...415L..27I,2017MNRAS.472.2608S,2021arXiv210803850I,2021MNRAS.508.2556I}. Because {\NeV} is a very-high ionization line with an ionization potential of 97.1 eV, its relative flux to {\Hb} can be used to constrain the contribution of very hard (i.e., X-ray) radiation to the ionizing spectrum shapes \citepe{2017AA...602A..45L,2020MNRAS.494..941S,2021arXiv210812438S}. \cite{2021arXiv210812438S} show that some ULX models can reproduce {\HeII}\W4686 and {\NeV}\W3426 flux values that are comparable to those from observations.}\par
\Add{We conduct the same method as described in \ref{sec:methods}, but add {\NeV}\W3426 into consideration. We only examine for {\EMPGb} because {\NeV} emission lines are not detected in the other two galaxies, most likely due to the lack of sensitivity. To avoid uncertainty in abundance ratio, we use {\NeV}\W3426/{\NeIII}\W3869 as we have done to {\SII}\W\W6316,6731. The observed {\NeV}\W3426/{\NeIII}\W3869 value and its 1$\sigma$ error for {\EMPGb} are $0.00347\pm0.00315$ (\citetalias{2021arXiv210512765B}). We obtain similar best-fit values for each parameter from MCMC. In Figure \ref{olivier}, we have overplotted the ionizing spectrum produced from the parameter in Table \ref{table:withneon} as the red dashed line. Although the  blackbody component has a higher temperature, the newly obtained ionizing spectrum generally has a similar spectrum shape as the one we obtain without considering {\NeV}\W3426/{\NeIII}\W3869. {\NeV}\W3426/{\NeIII}\W3869 value produced from new ionizing spectrum shapes is produced within around 3$\sigma$  of the observed value. Moreover, this new ionizing spectrum shape produces other emission lines within around 3$\sigma$. Therefore, we confirm that the general spectrum shape for {\EMPGb} is consistent with {\NeV}\W3426, which requires very high energy ($>97.1$ eV) ionizing photons.}

\input{neon_bp.tex}

\section{Discussion}\label{sec:discuss}
\subsection{Contributions from Stellar Radiation}
Our best-fit ionizing spectra have blackbody temperature around $3-5\times 10^4~{\rm K}$. B-type (O-type) stars have surface temperature of $1-3\times10^4~{\rm K}$ ($\geq3\times10^4~{\rm K}$). This suggests that blackbody radiation in our best-fit ionizing spectra, especially for that of {\EMPGb}, can be explained by B- and O-type stars. However, the convex downward spectra of {\EMPGa} and {\EMPGc} suggest that contributions from stars to hard photons around 35.1 to 54.4 eV might not be as significant as expected in BPASS binary models. Stellar models including fast-rotating stars \citepe{2019A&A...623A...8K} and population III stars \citepe{2021MNRAS.501.5517G} also have convex upward spectra around $13.6-54.4$ eV, unlike our results. These mismatches suggest that hard radiation from stellar sources cannot solely explain {\HeII} ionizing photons for the two of our galaxies with convex downward spectrum shapes.\par 
We also note that our general spectrum shapes do not take account of stellar absorptions. Stellar absorptions weaken intrinsic stellar radiation above ionization energies of ${\rm H^0}$ (13.6 eV), ${\rm He^0}$ (24.6 eV), and ${\rm He^+}$ (54.4 eV). This raises the question of whether stellar models with stellar absorption can really reproduce ionizing spectra of our three galaxies including the blackbody-dominated spectra of {\EMPGb}. More sophisticated spectral modeling is needed to further examine the stellar contributions, but this is beyond the scope of our paper.\par

\subsection{Contributions from Black Holes}
We examine the contributions from black holes (BHs) such as AGNs and HMXBs/ULXs. BHs can produce very hard radiation through non-thermal radiation from  Comptonization/synchrotron radiation. Because non-thermal radiation has a power-law-like shape, radiation from BH could explain the power-law-like component seen in the ionizing spectrum of {\EMPGa} and {\EMPGc}. BH radiation also has hot ($\gtrsim10^5~{\rm K}$) thermal radiation from their accretion disks. Temperatures of accretion disks are roughly proportional to $M_{\rm BH}^{-1/4}$, where $M_{\rm BH}$ is the mass of BH. Following the relation between the accretion disk temperature and $M_{\rm BH}$,
intermediate-mass BHs ($10^2\leq M_{\rm BH}\leq10^5~M_{\odot}$) and less massive ones would not produce sufficient ionizing 
photons in the soft energy range of $<54.4$ eV. For this reason, we will first examine if AGN can solely account for the general spectrum shapes of our three galaxies by comparing them with spectra of AGNs. Then, we will investigate whether the combinations of a BH and stellar population can explain the general spectrum shapes of our three galaxies.\par
\subsubsection{Only BHs}

\begin{figure}[htbp]
\includegraphics[width = \linewidth,clip]{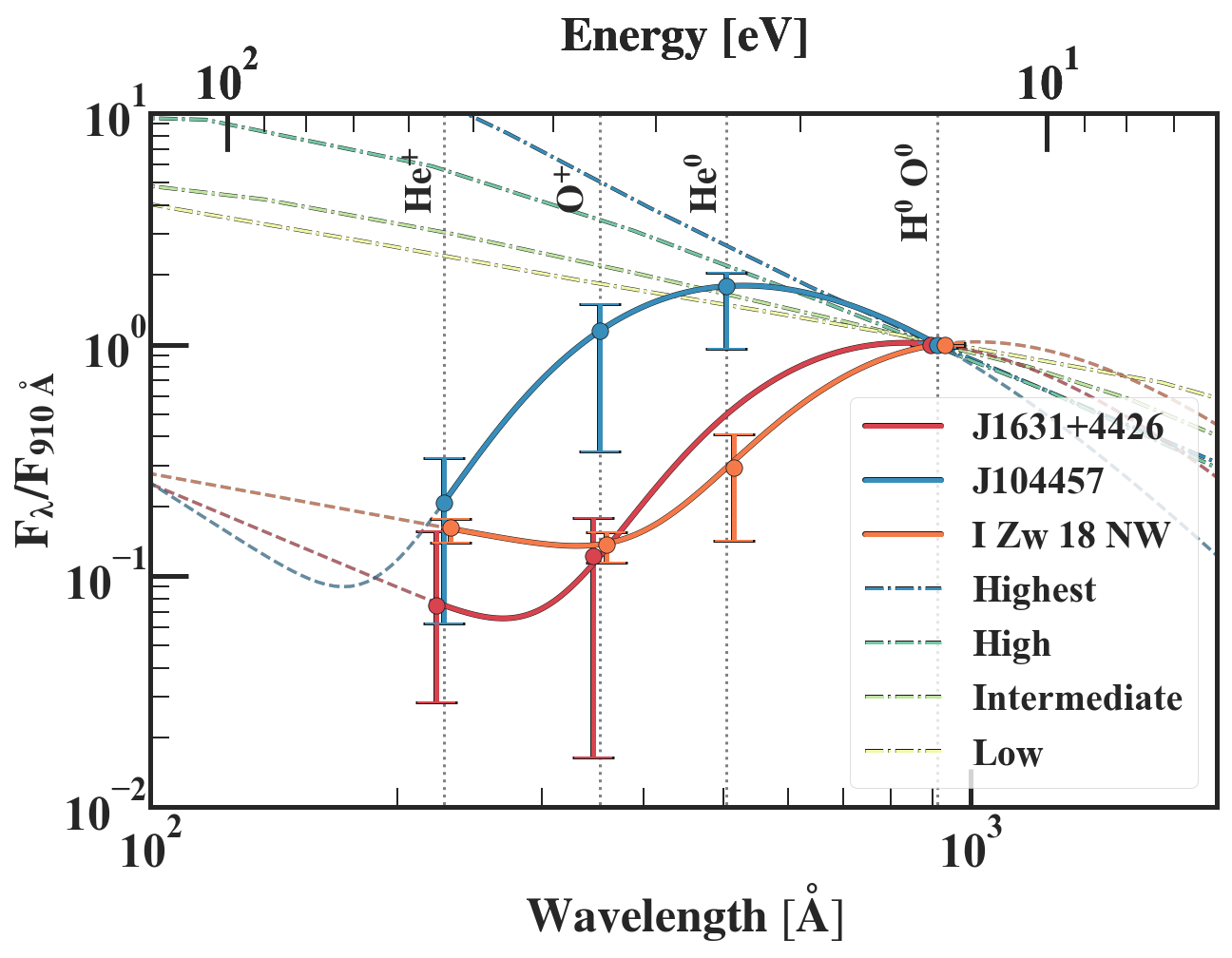}
\caption{Comparison of our result with AGN spectra. The dash-dotted lines represent mean AGN spectra by different Eddington ratios (Highest, High, Intermediate, and Low) taken from \cite{2020MNRAS.494.5917F}. The other symbols are the same as those in Figure \ref{seds}.}
\label{4agn}
\end{figure}

The inner-disk temperature $T_{\rm in}$ of the accretion disk is around $10^{4-5}~{\rm K}$ for AGNs. The blackbody radiation from the inner disk of an AGN gives a Big Blue Bump (BBB) feature in the UV range of the ionizing spectrum. The BBB feature combined with non-thermal radiation can produce a spectrum shape like the one reproduced from the combination of a blackbody and power-law radiation. We compare the shapes of four different AGN spectra with the general spectrum shapes of our three galaxies in Figure \ref{4agn}. These four AGN spectra are taken from \cite{2020MNRAS.494.5917F}, and they represent mean spectra of observed Type-1 AGNs grouped by Eddington ratios (Highest, High, Intermediate, and Low). As shown in Figure \ref{4agn}, all four AGN spectra have power-law like shapes between $13.6-54.4$ eV, unlike the ionizing spectra of our three galaxies. The deformation of disk-blackbody by the Compotonization in observed AGN might be the reason why AGN spectra does not have blackbody-like shapes in $13.6-54.4$ eV \citep[see][]{2020MNRAS.494.5917F}. From this comparison, we cannot conclude that AGN can solely explain the general spectrum shapes for our three galaxies. However, we note that ionizing spectrum shapes of AGNs remain unclear especially around the Lyman limit to soft X-ray because of the intervening absorption.\par

\subsubsection{BHs and Stars}
We will explore the combinations of BHs and stellar sources to see if they can account for the general spectrum shapes of our three galaxies. We choose ULXs as non-stellar sources because they are popular candidates for the origin of {\HeII}. Moreover, there are detections of a ULX in I Zw 18 through X-ray observations \citepe{2013ApJ...770...20K}. As a ULX spectrum, we use a spectral model described in \cite{2009MNRAS.392.1106G}. According to \cite{2021arXiv210812438S}, this ULX spectral model reproduces emission line ratios in metal-poor galaxies better than other ULX spectral models. \Add{We produce the ULX spectrum models from numerical calculations. To run the numerical calculations, we have referred to the code of the DISKIR model provided in {\sc xspec} \citep{1996ASPC..101...17A}.} As presented in \cite{2021arXiv210812438S}, ULX spectral models cannot produce enough photons in UV region, we combine ULX spectral model with stellar spectral model (hereafter ULX $+$ BPASS single model). We use spectra reproduced from BPASS single star models as stellar spectra. We assume single-aged stellar population with the same IMF as the BPASS binary model we presented in Figure \ref{seds}. We do not use BPASS binary models because adding extra hard components from ULX would not alter the convex upward spectrum shape of the spectra reproduced from BPASS binary models. We use fluxes at the ionization energies of ${\rm H^0}$, ${\rm He^0}$, ${\rm O^+}$, and ${\rm He^+}$ for the best-fit ionizing spectra to restrict the ULX $+$ BPASS single models for our three galaxies. We allow a stellar age $t$ and a ratio between $0.5-8$ keV X-ray luminosity $L_{\rm X}$ and SFR ($L_{\rm X}/{\rm SFR}$) to vary. A SFR for $L_{\rm X}/{\rm SFR}$ is calculated from UV luminosity at 1500 {\AA} using relationships presented in \cite{1998ARA&A..36..189K}. \Add{For each of our three galaxies, we fix the value of stellar metallicity $Z_*$ at the best-fit value of gas-phase metallicity.}\par

\input{ulbp_params1.tex}

\begin{figure}[htbp]
\includegraphics[width = \linewidth,clip]{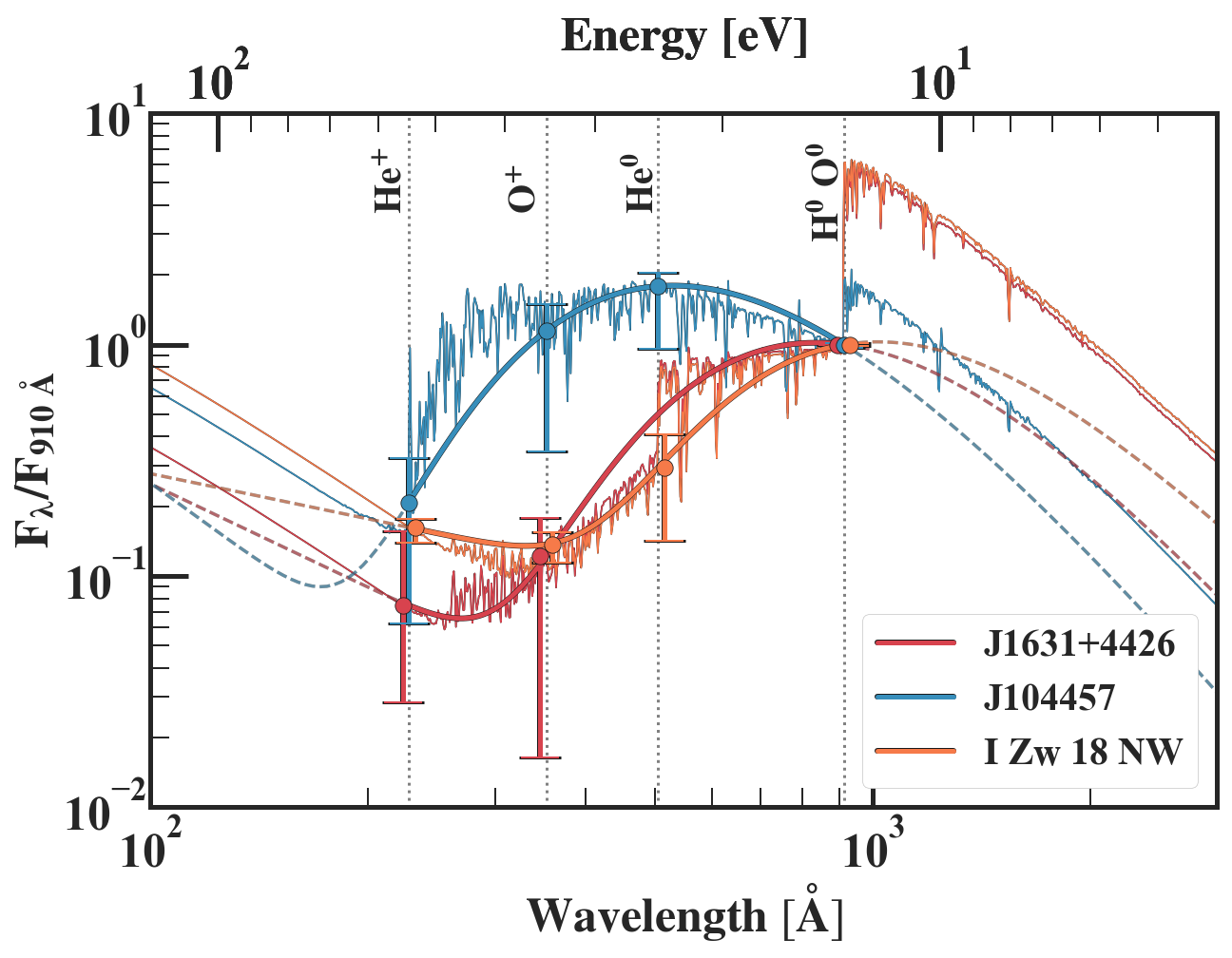}
\caption{ULX+BPASS single models for {\EMPGa} (red thin line), {\EMPGb} (blue thin line), and {\EMPGc} (orange thin line). The other symbols are the same as those in Figure \ref{seds}.}
\label{ulbp}
\end{figure}

In Table \ref{table:ulbp_params}, we summarize the parameters of the ULX + BPASS single models that reproduce the general spectrum shapes for our three galaxies. We compare the general spectrum shapes with the ULX + BPASS single star models in Figure \ref{ulbp}. The spectrum of ULX + BPASS single star models for {\EMPGa}, {\EMPGb}, and {\EMPGc} are represented by the red, blue, and orange thin lines, respectively. We successfully reproduce the general spectrum shape of the best-fit ionizing spectra between $13.6-54.4$ eV for our three galaxies, despite the diversity in the general spectrum shapes. For both types of the spectrum shapes, the $L_{\rm X}/{\rm SFR}$ values are $\lesssim10^{41}~{\rm erg s^{-1}/M_\odot {yr}^{-1}}$, which are comparable to the values obtained in \cite{2021arXiv210812438S}. The blackbody-like spectrum shape of {\EMPGb} can be explained by prominent stellar contributions relative to that of the ULX, while the convex downward spectrum shapes of {\EMPGa} and {\EMPGc} can be explained by the less prominent stellar contributions around $35.1-54.4$ eV than that of {\EMPGb}. Moreover, the stellar component for {\EMPGb} has the stellar age around 2 Myr, while the stellar components for {\EMPGa} and {\EMPGc} have the stellar age above 5 Myr. This suggests that the diversity of the spectrum shapes can arises from the stellar age difference of the stellar components. The stellar contributions become less prominent at $\gtrsim$ 5 Myr because massive stars with their mass at $\gtrsim40~M_\odot$ already reach their lifetimes \citep{2018MNRAS.477..904X}. \par
\subsection{Other Possibilities}
We have only examined ULX as the non-stellar component when considering the combination with the stellar component, but it does not limit the possibilities of other sources such as low luminosity AGNs and supersoft X-ray sources. Photoionization by shocks is still not ruled out from the candidate by our analysis. \Add{The ionization of {\HeII} by soft X-rays from cluster winds and superbubbles also remains a possibility \citep[see][]{2022arXiv220304987O}.} Further X-ray observations for {\EMPGa} and {\EMPGb} may enable us to constrain contributions from X-ray emitting sources.\par

\subsection{Multi-Wavelength Comparison and Relation with Cosmic Reionization}

\begin{figure}[htbp]
\includegraphics[width = \linewidth,clip]{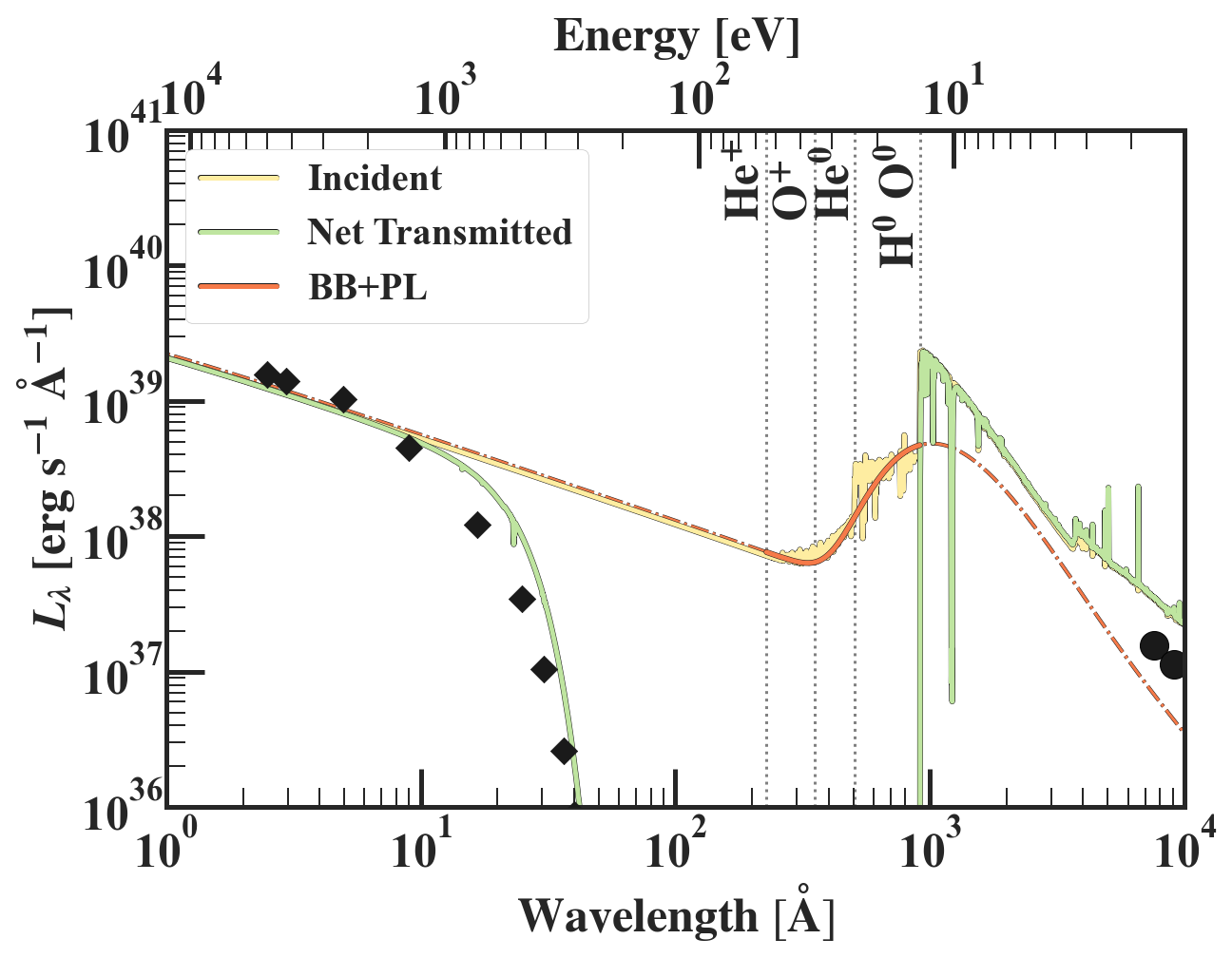}
\caption{Comparison between the multi-wavelength data and our models for I Zw 18. The black diamonds represent the unfolded X-ray luminosities. The leftward (rightward) black circle represents the $i'$- ($z'$-) band magnitudes. The yellow (green) line represents the ionizing spectrum (net transmitted spectrum) of the comparison model. We also overplot the ionizing spectrum of I Zw 18NW in Figure \ref{seds} (BB+PL, orange line) under an assumption that the whole system has the same spectral shape. The other symbols are the same as those in Figure \ref{seds}.}
\label{t_compare}
\end{figure}

We investigate the consistency between our ionizing spectrum shapes and the observed values of I Zw 18, where multi-wavelength data from the X-ray to optical photometry data are available. We assume that whole I Zw 18 has the same ionizing spectrum shape as our result for {\EMPGc} to compare the model with the observed values of I Zw 18 as a whole. We use the X-ray luminosity obtained by the \textit{X-ray Multi-Mirror Mission} (\textit{XMM-Newton}) \citep{2013ApJ...770...20K,2017AA...602A..45L}. For optical photometry, we use SDSS $i'$- and $z'$-band magnitudes from DR16 \citep{2020ApJS..249....3A} for I Zw 18 NW and I Zw 18 SE to compare the fluxes of the continuum. We do not use photometry data in other bands, where many strong emission lines are detected. To compare the optical photometry data, we replace the blackbody component of the ionizing spectrum with the 
BPASS single star model reproduced with the parameters for {\EMPGc} in Table \ref{table:ulbp_params}. We normalize the luminosity by adopting a distance of 18.2 Mpc \citep{2007ApJ...667L.151A}, {\Hb} luminosity of $10^{39.43}$ {\ergs}, and a relation between the number of hydrogen ionizing photons produced per second and the {\Hb} luminosity presented in \cite{2010ApJ...724.1524O}. We show the comparison between the multi-wavelength data and the comparison model in Figure \ref{t_compare}. The black diamonds represent unfolded X-ray luminosities and the leftward (rightward) black circle represents the $i'$- ($z'$-) band magnitude. The incident (net transmitted) spectrum of the comparison model is shown in the yellow (green) line. Here, the net transmitted spectrum is the sum of attenuated incident radiation and the outward portion of the emitted continuum and lines. The orange line is the same ionizing spectrum for {\EMPGc} as the one presented in Figure \ref{seds}. From Figure \ref{t_compare}, we find that the optical photometry data is slightly below what our comparison model predicts. However, the net transmitted spectrum of our comparison model is consistent with the X-ray luminosities.\par

\begin{figure}[htbp]
\includegraphics[width = \linewidth,clip]{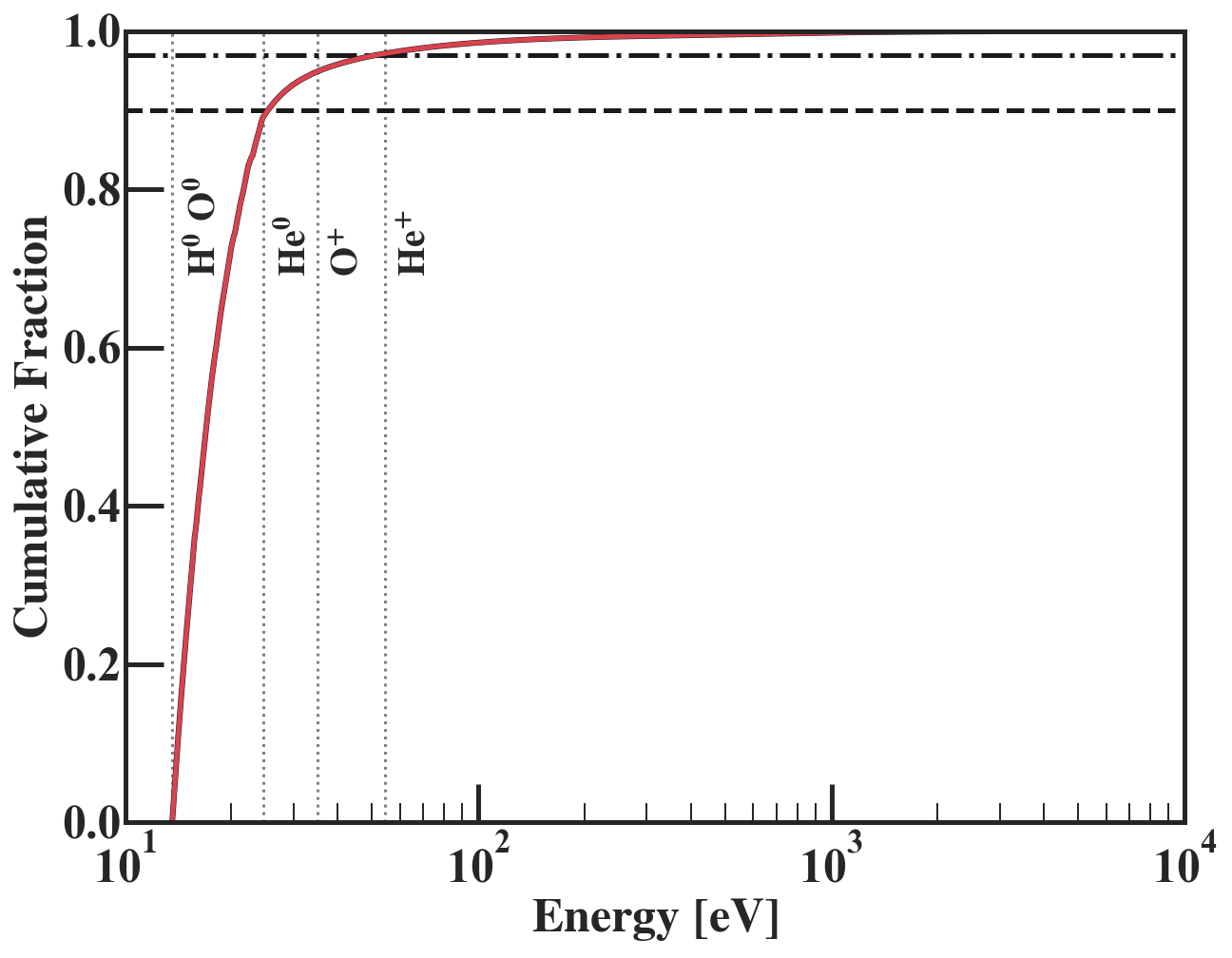}
\caption{Cumulative fraction of the hydrogen ionizing photon numbers at energies (red line) above the ionization energy of ${\rm H^0}$ (13.6 eV) for our spectrum model. The dotted lines mark the ionization energies of ${\rm H}^{0}$, ${\rm O}^{0}$. ${\rm He}^{0}$, ${\rm O}^{+}$, and ${\rm He}^{+}$. The cumulative fraction of the ionizing photon numbers at the ionization energy of ${\rm He^{0}}$ (24.6 eV) already reaches $\sim90\%$ (dashed line). The cumulative fraction at the ionization energy of ${\rm He^{+}}$ (54.4 eV) exceeds $97\%$ (dotted-dashed line).}
\label{ion_frac}
\end{figure}

As described in Section \ref{sec:intro}, a nearby EMPG can be a probe for young galaxies at EoR. We evaluate the contribution of X-ray photons from the young galaxies to cosmic reionization, assuming that the young galaxies have the same ionizing spectrum shapes as the spectrum of I Zw 18. We adjust the ionizing photon escape fraction of 20\% to match the standard picture of cosmic reionization \citepe{2009ApJ...706.1136O,2013ApJ...768...71R}. To do so, we consider the model spectrum that linearly combines the incident ionizing spectrum and the net transmitted spectrum with their luminosities multiplied by 0.2 and 0.8, respectively. In Figure \ref{ion_frac}, we depict the cumulative fraction of the hydrogen ionizing photon number at each energy in the red line. As shown in Figure \ref{ion_frac}, about $90\%$ ($97\%$) of the ionizing photons have energy below 24.6 eV (54.4 eV). Moreover, the contributions of the high energy ($>$54.4 eV) photons would be further reduced because the photoionization cross-section has the photon energy dependency of approximately $E^{-3}$ \citep{2006agna.book.....O}. Thus, we conclude that the hard ionizing radiation does not affect the standard picture of cosmic reionization \citep{2021arXiv211013160R}.\par

\section{Conclusions}\label{sec:conc}

We have searched for the general spectrum shapes of ionizing sources in EMPGs that can explain major emission lines including {\HeII}. We have prepared parameterized ionizing spectra with generalized shapes, and use these ionizing spectra in photoionization models with free parameters of ionizing spectra and nebulae. For three EMPGs with strong {\HeII\W4686}, we have determined seven parameters of the ionizing spectra and nebulae using MCMC methods.\par
Our main findings can be summarized as follows:
\begin{enumerate}
   \item We have determined the best-fit parameters for our three galaxies. All the best-fit parameters are comparable to ordinary values. With the best-fit parameters, we successfully reproduced all major emission lines within $\lesssim3\sigma$ errors. 
   \item The ionizing spectrum produced from the best-fit parameters for {\EMPGb} has a blackbody-dominated shape around $13.6-54.4$ eV with blackbody temperature $5\times10^4$ K. In contrast, the ionizing spectra for {\EMPGa} and {\EMPGc} have significant contributions from power-law radiation around 54.4 eV, with relatively low-temperature ($3\times10^4$ K) blackbody radiation. As a result, the general spectrum shapes for {\EMPGa} and {\EMPGc} have convex downward shapes around $13.6-54.4$ eV, which are fundamentally different from the shapes predicted by BPASS binary models. These two types of spectrum shapes indicate a diversity in ionizing spectrum shapes in EMPGs.
   \item We compare obtained general shapes of ionizing spectra with that of ionizing spectra from stars, AGNs, and ULXs. We find that ULX $+$ BPASS single star models can explain the general shapes of the obtained ionizing spectra for our three galaxies. Furthermore, we show that the diversity of the spectrum shapes can be attributed to the differences in the stellar age.
   \item We confirm that our result is consistent with the X-ray to optical photometry data for I Zw 18, of which the multi-wavelength data are available. We also find that the hard ionizing radiation inferred from our general spectrum shape of I Zw 18 does not contradict the standard picture of cosmic reionization.
\end{enumerate}
\par
\Add{Spectroscopic observation of {\EMPGa} and {\EMPGc} for optical {\NeV} emission lines with high sensitivity would also be helpful in terms of constraining contributions of ionizing photons $\gtrsim100$ eV in {\HeII} emissions. Our new analysis method would provide new insights into the statistical nature of EMPGs if applied to the sample with a large number of EMPGs.}

\section*{Acknowledgements}\label{sec:ack}
We thank Wako Aoki, Yoshihisa Asada, Kohei Inayoshi, Akio Inoue, \Add{Carolina Kehrig}, Tomoya Kinugawa, Haruka Kusakabe, Vianney Lebouteiller, Daniel Schaerer, Yuta Tarumi, and Masayuki Umemura for sharing their data and figures and giving us helpful comments. This research is based in part on data collected at Subaru Telescope, which is operated by the National Astronomical Observatory of Japan. We are honored and grateful for the opportunity of observing the Universe from Maunakea, which has the cultural, historical and natural significance in Hawaii. This work is supported by the World Premier International Research Center Initiative (WPI Initiative), MEXT, Japan, as well as KAKENHI Grant-in-Aid for Scientific Research (A) (20H00180, and 21H04467) through the Japan Society for the Promotion of Science (JSPS). This work was supported by the joint research program of the Institute for Cosmic Ray Research (ICRR), University of Tokyo. Numerical computations were in part carried out on Cray XC50 at Center for Computational Astrophysics, National Astronomical Observatory of Japan.

\bibliography{paper.bib}{}
\bibliographystyle{aasjournal}



\end{document}

%% file: galaxy.tex
\begin{deluxetable*}{cccccccccc}
\tablecolumns{10}
\tabletypesize{\scriptsize}
\tablecaption{Sample Properties%
\label{table:gal}}
\tablehead{%
\colhead{Name of the Galaxy} & 
\colhead{$z$} &
\colhead{$i$} &
\colhead{$\log$($M_{\star}$)} &
\colhead{$\log$(SFR)} &
\colhead{$Z$} &
\colhead{$F$(H$\beta$)} &
\colhead{{\EW}(H$\beta$)} &
\colhead{$L_{{\rm X,}0.5-10~{\rm keV}}$} &
\colhead{References}\\
\colhead{} &
\colhead{} &
\colhead{(mag)} &
\colhead{($M_{\odot}$)} &
\colhead{({$M_{\odot}$\,yr$^{-1}$})} &
\colhead{($Z_{\odot}$)} &
\colhead{($10^{-14}~$\ergscm)} & 
\colhead{(\AA)} &
\colhead{($10^{39}$~\ergs)} &
\colhead{}\\
\colhead{} &
\colhead{(1)} &
\colhead{(2)} &
\colhead{(3)} &
\colhead{(4)} &
\colhead{(5)} &
\colhead{(6)} &
\colhead{(7)} & 
\colhead{(8)} &
\colhead{(9)}
}
\startdata
{\EMPGa}  &  \comment{$0.03125$}$0.031$  & $22.5$ & $5.9$\comment{^{+0.10}_{-0.09}} & $-1.28$\comment{\pm0.01} & $0.016$ & $0.13$\comment{\pm0.04} & $123.5$\comment{^{+3.5}_{-2.8}} & \nodata & (a)\\ 
{\EMPGb}  &  $0.013$  & $18.7$\comment{\pm0.02}  & $6.8$ & $-0.85$ & $0.058$ & $0.95$\comment{\pm0.10} & \nodata & \nodata & (b), (c)\\
{\EMPGc}  & \comment{$0.00243$}0.002  & $20.1$\comment{\pm0.10} & $5.5$ & $-2.70$ & $0.030$ & $3.19$ & $81.2$ & $1.4-1.6$ & (d), (e), (f), (g)\\
\enddata
\tablecomments{
(1): Redshift.
(2): $i$-band magnitude. The magnitudes are the {\tt cmodel} magnitudes of Subaru/HSC (SDSS)
for {\EMPGa} ({\EMPGb} and {\EMPGc}).
(3): Stellar mass.
(4): Star-formation rate (SFR). The SFR of {\EMPGc} is derived from the {\Ha} flux in the same manner as \cite{2021ApJ...918...54I}.
(5): Gas-phase Metallicity derived by the direct $T_{\rm e}$ method.
(6): {\Hb} flux.
(7): Rest-frame equivalent width of an {\Hb} emission line.
(8): X-ray luminosity at $0.5-10.0$ keV obtained by ${\it Chandra}$ X-ray observations.
(9): References for values in the columns $(1)-(8)$: (a) \cite{2020ApJ...898..142K}, (b) \citetalias{2021arXiv210512765B}, (c) SDSS DR2 \citep{2004AJ....128..502A}, (d) SDSS DR6 \citep{2008ApJS..175..297A}, (e) \citetalias{2005ApJS..161..240T}, (f) \cite{2017AA...602A..45L}, (g) \cite{2004ApJ...606..213T}
}
\end{deluxetable*}

%% file: flux1.tex
\begin{deluxetable}{lccc}
\tablewidth{\textwidth}
\setlength{\tabcolsep}{8pt}
\tabletypesize{\scriptsize}
\tablecaption{Fluxes and Flux ratios}
\tablehead{{Ion}       & {\EMPGa}         & {\EMPGb} & {\EMPGc} \vspace{+0.05cm}}
\startdata
\multicolumn{4}{c}{{\boldmath$F_{\lambda}/F_{{\rm H\beta}}$}} \vspace{+0.05cm} \\
\hline \vspace{-0.15cm} \\
{\OII}\W\W3727,3729       & 50.12$\pm$2.66          & 26.37$\pm$0.27 & 37.66$\pm$0.61               \\
{\HeI}\W4026        & \nodata           & 1.72$\pm$0.03 & 1.03$\pm$0.09           \\
{\Hd}        & 27.53$\pm$0.65           & 25.59$\pm$0.42 & 26.52$\pm$0.42           \\
{\Hc}   & 46.88$\pm$0.50           & 46.55$\pm$0.67 & 47.53$\pm$0.71           \\
{\OIII}\W4363        & 8.18$\pm$0.48          & 13.51$\pm$0.21 & 6.74$\pm$0.14           \\
{\HeI}\W4471        & \nodata          & 3.97$\pm$0.09 & 2.76$\pm$0.08           \\
{\HeII}\W4686        & 2.32$\pm$0.38           & 1.80$\pm$0.04 & 3.44$\pm$0.09           \\
{\Hb}        & 100.00$\pm$0.37                & 100.0$\pm$1.4 & 100.00$\pm$1.45           \\
{\OIII}\W4959        & 55.76$\pm$0.34                & 143.0$\pm$1.5 & 67.15$\pm$0.97           \\
{\OIII}\W5007        & 170.92$\pm$0.38           & 427.5$\pm$4.3 & 200.62$\pm$2.89           \\
{\Ha}        & \nodata          & 296.7$\pm$4.2 & \nodata  \\
\vspace{-0.1cm} \\
\hline \vspace{-0.4cm} \\
\multicolumn{4}{c}{\bf Line Ratios} \vspace{+0.05cm} \\
\hline 
\vspace{-0.1cm} \\
{\SII}\W6316/\W6731             & $1.41\pm0.40^a$ & 1.23$\pm$0.04   & 1.30$\pm$0.61   \\
\vspace{-0.1cm} \\
\enddata
\tablecomments{Dust-corrected emission line fluxes and emission line ratios for all galaxies in our sample. 
The fluxes are all scaled with {\Hb}$=100$, and taken from \citetalias{2021ApJ...913...22K}, \citetalias{2021arXiv210512765B}, and \citetalias{2005ApJS..161..240T}, except for {\SII} \W6316/\W6731 of {\EMPGa}. \tablenotetext{a}{This line ratio of {\EMPGa} is measured with the spectrum of \cite{2020ApJ...898..142K} in the same manner as \citetalias{2021ApJ...913...22K}.}}
\label{table:flux1}
\end{deluxetable}

%% file: prior.tex
\begin{deluxetable}{lcc}
\tablewidth{\textwidth}
\tablecolumns{3}
\tabletypesize{\scriptsize}
\tablecaption{Prior Distributions for Free Parameters
\label{table:prior}}
\tablehead{{} & \hspace{1.5cm}{Parameter}\hspace{1.5cm} & \hspace{1.5cm}{Prior Range}\hspace{.51cm}}
\startdata
(1) & $\log a_{\rm mix}$ & $[-3,3]$\\
(2) & $\log T_{\rm BB}/{\rm K}$ & $[4,6]$ \\
(3) & $\alpha_{\rm X}$ & $[-3,1]$ \\ 
(4) & $\log U$ & $[-5,-0.5]$ \\
(5) & $\log n_{\rm H}/{\rm cm^{-3}}$ & $[0.5,5]$ \\
(6) & $\log Z/Z_\odot$ & $[-3,0]$ \\
(7) & $\mathit{N}_{{\rm H}\beta}$ & within $3\sigma_{{\rm H}\beta,{\rm obs}}$
\enddata
\tablecomments{
(1): Ratio of the blackbody flux to the power-law flux at 1 Ryd.
(2): Blackbody temperature. 
(3): Power law index.
(4): Ionization parameter.
(5): Hydrogen density.
(6): Gas-phase metallicity.
(7): Normalization factor for an {\Hb} emission line \Add{(i.e., model emission line flux for {\Hb})}. 
\Add{Here, the observed {\Hb} emission line flux values are normalized at 100.}}
\end{deluxetable}

%% file: best_params.tex
\begin{deluxetable}{ccccc}
\tablewidth{\textwidth}
\tablecolumns{5}
\tabletypesize{\scriptsize}
\tablecaption{Best Fit Parameters
\label{table:best_params}}
\tablehead{ {} & {Parameters}       & {\EMPGa}         & {\EMPGb} & {\EMPGc} \vspace{+0.05cm}}
\startdata
\multicolumn{5}{c}{\bf Best Fit Parameters} \vspace{+0.05cm} \\
\hline \vspace{-0.15cm} \\
(1) & $\log a_{\rm mix}$ & $-2.05^{+0.74}_{-0.84}$ & $-2.81^{+0.83}_{-0.08}$ & $-1.18^{+0.14}_{-0.35}$ \\
(2) & $\log T_{\rm BB}/{\rm K}$ & $4.54^{+0.02}_{-0.14}$ & $4.74^{+0.03}_{-0.11}$ & $4.45^{+0.05}_{-0.12}$ \\
(3) & $\alpha_{\rm X}$ & $-0.53^{+0.70}_{-0.71}$ & $0.26^{+0.09}_{-0.80}$ & $-1.37^{+0.45}_{-0.20}$ \\
(4) & $\log U$ & $-1.55^{+0.05}_{-0.68}$ & $-1.95^{+0.34}_{-0.07}$ & $-1.67^{+0.22}_{-0.34}$ \\
(5) & $\log n_{\rm H}/{\rm cm^{-3}}$ & $1.74^{+2.48}_{-0.39}$ & $2.82^{+0.19}_{-0.63}$ & $2.12^{+0.97}_{-0.82}$ \\
(6) & $\log Z/Z_\odot$ & $-1.61^{+0.19}_{-0.05}$ & $-1.19^{+0.05}_{-0.03}$ & $-1.48^{+0.07}_{-0.05}$ \\
(7) & $\mathit{N}_{{\rm H}\beta}$ & $99.9^{+0.9}_{-0.4}$ & $102.3^{+1.9}_{-3.5}$ & $101.9^{+1.4}_{-1.1}$  \\
\enddata
\tablecomments{Best fit parameters are the sampled parameters that maximize $\log\mathcal{L}$ in equation (\ref{logL}). Uncertainties for best fit parameters are derived from condition shown in equation (\ref{uncert}).
(1): Ratio of blackbody flux to power-law flux at 1 Ryd.
(2): Blackbody temperature. 
(3): Power law index.
(4): Ionization parameter.
(5): Hydrogen density.
(6): Gas-phase metallicity.
(7): Normalization factor for an {\Hb} emission line. \Add{(i.e., model emission line flux for {\Hb})}. 
\Add{Here, the observed {\Hb} emission line flux values are normalized at 100.}
}
\end{deluxetable}

%% file: neon_bp.tex
\begin{deluxetable}{lcc}
\tablewidth{\textwidth}
\tablecolumns{3}
\tabletypesize{\scriptsize}
\tablecaption{\Add{Best Fit Parameters with {\NeV} for {\EMPGb}}
\label{table:withneon}}
\tablehead{{} & \hspace{1.5cm}{Parameters}\hspace{1.5cm} & \hspace{1.5cm}{Best Fit Values}\hspace{.51cm}}
\startdata
(1) & $\log a_{\rm mix}$ & $-2.97$\\
(2) & $\log T_{\rm BB}/{\rm K}$ & $4.80$ \\
(3) & $\alpha_{\rm X}$ & $0.37$ \\ 
(4) & $\log U$ & $-2.05$ \\
(5) & $\log n_{\rm H}/{\rm cm^{-3}}$ & $2.84$ \\
(6) & $\log Z/Z_\odot$ & $-1.16$ \\
(7) & $\mathit{N}_{{\rm H}\beta}$ & $102.8$
\enddata
\tablecomments{
(1): Ratio of the blackbody flux to the power-law flux at 1 Ryd.
(2): Blackbody temperature. 
(3): Power law index.
(4): Ionization parameter.
(5): Hydrogen density.
(6): Gas-phase metallicity.
(7): Normalization factor for an {\Hb} emission line.
\Add{The best fit values listed in this table are obtained with the analysis described in Section \ref{sec:methods} but with taking {\NeV}\W3462/{\NeIII}\W3869 into consideration.}
}
\end{deluxetable}

%% file: ulbp_params1.tex
\begin{deluxetable}{ccc}
\tablewidth{\linewidth}
\tablecolumns{3}
\tabletypesize{\scriptsize}
\tablecaption{Parameters for ULX + BPASS Single Models
\label{table:ulbp_params}}
\tablehead{\colhead{Name of the Galaxy} & 
\colhead{$t$} & 
\colhead{$\log L_{{\rm X}}/{\rm SFR}$} \\
\colhead{} &
\colhead{(Myr)} &
\colhead{({\ergs}/$M_{\odot}\,{\rm yr}^{-1}$)} \\
\colhead{} &
\colhead{(1)} &
\colhead{(2)} 
}
\startdata
{\EMPGa} & $8.2$ & $40.7$ \\
{\EMPGb} & $2.2$ & $41.5$ \\
{\EMPGc} & $8.5$ & $41.0$ \\
\enddata
\tablecomments{
\Add{(1): Stellar Age. We assume a single-aged stellar populations.
(2): $0.5-8$ keV X-ray luminosity to SFR.}
}
\end{deluxetable}